\pdfoutput=1
\documentclass[conference]{IEEEtran}
\IEEEoverridecommandlockouts

\usepackage{cite}
\usepackage{amsmath,amssymb,amsfonts}
\usepackage{booktabs} 
\usepackage{textcomp}
\usepackage{xcolor}
\usepackage{multirow}
\usepackage{amsmath}
\usepackage{graphicx}
\usepackage{pifont}
\usepackage{subfig}
\usepackage[noend]{algorithmic}
\usepackage{url}
\usepackage{algorithm}
\usepackage{listings}
\usepackage{color}
\usepackage{fancyhdr}
\usepackage{comment}

\definecolor{dkgreen}{rgb}{0,0.6,0}
\definecolor{gray}{rgb}{0.5,0.5,0.5}
\definecolor{mauve}{rgb}{0.58,0,0.82}

\usepackage{todonotes}

\newcommand{\yellowcomment}[1]{\todo[inline,color=yellow!40,size=\small]{#1}}

\def\BibTeX{{\rm B\kern-.05em{\sc i\kern-.025em b}\kern-.08em
    T\kern-.1667em\lower.7ex\hbox{E}\kern-.125emX}}
\begin{document}

%

\title{\Large \bf Towards QoS-Aware and Resource-Efficient GPU Microservices Based on Spatial Multitasking GPUs In Datacenters
} 
%

\author{

\IEEEauthorblockN{1\textsuperscript{st} Wei Zhang}
\IEEEauthorblockA{
	\textit{Shanghai Jiao Tong University}\\
	}

\and

\IEEEauthorblockN{2\textsuperscript{nd} Quan Chen}
\IEEEauthorblockA{
	\textit{Shanghai Jiao Tong University}\\
	}

\and
\IEEEauthorblockN{3\textsuperscript{rd}Kaihua Fu}
\IEEEauthorblockA{
	\textit{Shanghai Jiao Tong University}\\
	}

\and
\IEEEauthorblockN{4\textsuperscript{th} Ningxin Zheng}
\IEEEauthorblockA{
	\textit{Shanghai Jiao Tong University}\\
	}

\and

\IEEEauthorblockN{5\textsuperscript{th} Zhiyi Huang}
\IEEEauthorblockA{
	\textit{University of Otago}\\
	}
\and
\IEEEauthorblockN{6\textsuperscript{th} Jingwen Leng}
\IEEEauthorblockA{
	\textit{Shanghai Jiao Tong University}\\
	}

\and
\IEEEauthorblockN{7\textsuperscript{th} Chao Li}
\IEEEauthorblockA{
	\textit{Shanghai Jiao Tong University}\\
	}

\and
\IEEEauthorblockN{8\textsuperscript{th} Wenli Zheng}
\IEEEauthorblockA{
	\textit{Shanghai Jiao Tong University}\\
	}

\and
\IEEEauthorblockN{\centerline{9\textsuperscript{th} Minyi Guo}}
\IEEEauthorblockA{
	\textit{Shanghai Jiao Tong University}\\
	}

	
}

\maketitle

%

%
\begin{abstract}
While prior researches focus
on CPU-based microservices, they
are not applicable for GPU-based microservices due to the different contention patterns. It is 
challenging to optimize the resource utilization while
guaranteeing the QoS for GPU microservices. We find that the
overhead is caused by {\it inter microservice communication}, {\it GPU
resource contention} and {\it imbalanced throughput within
microservice pipeline}.
We propose Camelot, a runtime system that 
manages GPU micorservices considering the above factors.
In Camelot, a {\it global memory-based communication mechanism} 
enables onsite data sharing that significantly reduces the end-to-end latencies of user queries.
We also propose two {\it contention aware resource allocation policies}
that either maximize the peak supported service load or minimize the 
resource usage at low load while ensuring the required QoS. 
The two policies consider the microservice pipeline effect and the runtime
GPU resource contention when allocating resources for the microservices. 
Compared with state-of-the-art work, Camelot increases the supported peak load by 
up to 64.5\% with limited GPUs, and reduces 35\% resource usage at low load while achieving the desired 99\%-ile latency target. 
\end{abstract}

%
%

%

%

%

\section{Introduction} \label{sec:introduction}
Datacenters~\cite{barroso2009datacenter} host latency critical user-facing applications, such as web search~\cite{websearch} and web service~\cite{webservice}. 
These applications have strict Quality of Service (QoS) requirement in terms of tail latency, and require frequent bug fixing and feature updating. 
To meet these requirements, service design shifts from a monolithic architecture to a microservice architecture~\cite{gan2019open}, where a complex user-facing service is decomposed into multiple loosely coupled microservices. Each microservice provides a specialized functionality. {\color{black}A microservice-based  application involves the interoperation of multiple microservices, each of which can be implemented, deployed, and updated independently without compromising the application's integrity. Such independence improves the application's scalability, portability, and availability. Considering these advantages, the microservice architecture has been regarded as the widely accepted and employed software architecture by Internet giants such as Netflix, Amazon, Apple and eBay~\cite{li2019dataflow,evolution,workshop}.}

\begin{figure}
\centering
	\includegraphics[width=.9\columnwidth]{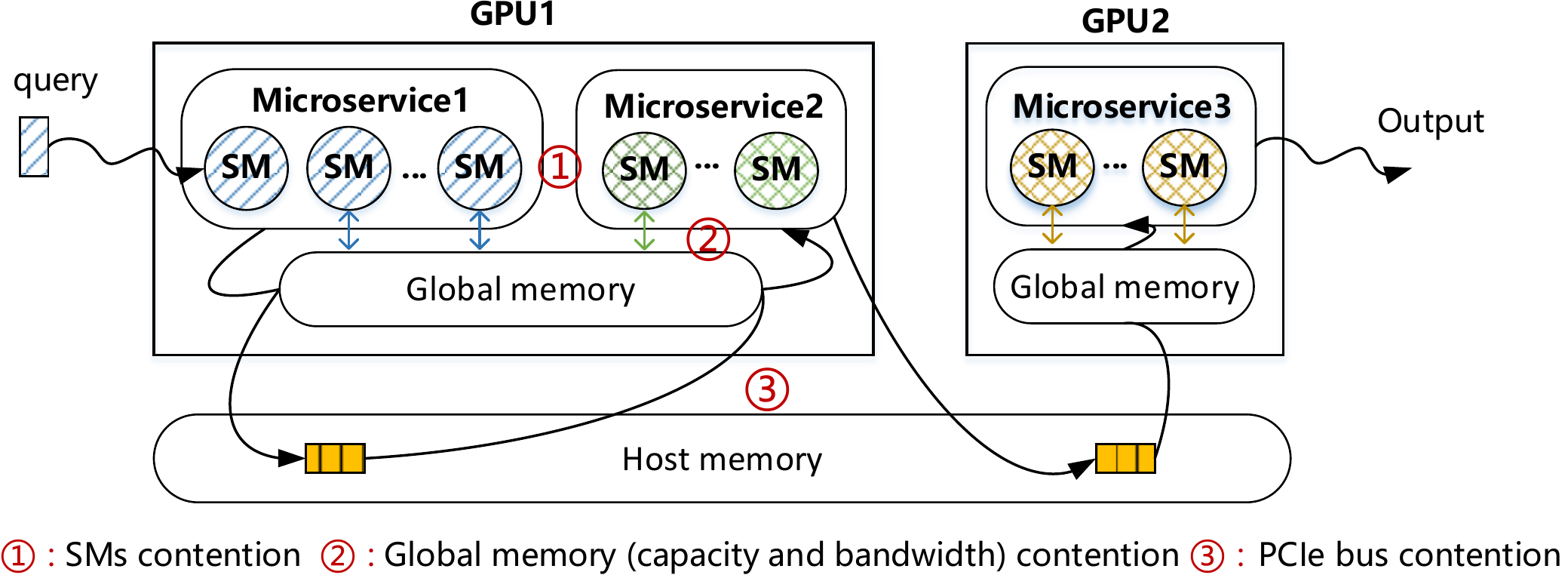}
	\vspace{-2mm}
	\caption{\label{fig:GPU-interference} An example of deploying GPU microservices.}
	\vspace{-6mm}
\end{figure}

Similarly, user-facing services on GPU (e.g., intelligent personal assistant~\cite{chen2016baymax}, graph processing~\cite{graph}, and deep learning~\cite{dl}) are also shifting towards the microservice architecture (referred as ``GPU microservices''). 
 Figure~\ref{fig:GPU-interference} shows an example of deploying an application that has three microservice stages on GPU. In the figure, multiple microservices run on a spatial multitasking GPU concurrently, since the current Volta Multi-Process Service(MPS)~\cite{voltamps} allows multiple applications to share GPU computational resources for better resource efficiency.
Observed from this figure, the back pressure effects caused by the dependencies between the microservices result in expensive overhead~\cite{gan2019open}. The cascading QoS violations will quickly propagate through the entire service, which leads to worse consequences of QoS violations. Therefore, even though the quality-of-service (QoS) requirements of user-facing applications are similar for microservices and monoliths, the tail latency required for each individual microservice is much stricter than for traditional monolith applications. 

Besides guranteeing the QoS of microservices, it is cost efficient to maximize the supported peak load of a user-facing application with limited resources, and minimize resource usage of a service with varying load.  
There are some prior researches on characterizing and managing resources for CPU microservices~\cite{sriraman2019softsku,bao2019performance,gan2019seer,li2019dataflow}. {\color{black}Benefit from the containerized  
deployment pattern~\cite{bao2019performance} of CPU microservices, such interference could be encapsulated and resolved at the container level. A container may be imposed with certain limits on the CPU and memory resources consumed by a microservice.} 

However, prior analysis and resource management policies do not apply for GPU microservices. While CPU microservices contend for CPU and memory bandwidth, 
GPU microservices contend for SMs, global memory capacity and bandwidth, and PCI-e bandwidth (as shown in Figure~\ref{fig:GPU-interference}). In addition, there is no containerized environment that enables fine-grained resource sharing for spatial multitasking GPUs. 
Balancing the throughput of the microservices to improve the microservice pipeline efficiency and eliminate the backpressure effect is challenging on GPU.
{\color{black}Laius~\cite{zhang2019laius} is state-of-the-art work that manages resource on spatial multitasking GPUs. It improves the GPU utilization by co-locating user-facing applications and batch applications when ensuring the QoS of the user-facing applications. However, Laius is not able to handle GPU microservices that show complex dependency relationship, because it assumes independent relationship between that the co-located tasks.}

We find that the {\it communication overhead} between 
microservices, the {\it pipeline efficiency} (determined by the number of SMs
allocated to each microservice, and the number of instances in each
microservice stage), and the {\it global memory bandwidth contention} together
determine the tail latencies of GPU microservices. 
{\color{black} We have two insights: 1) the communication between GPU microservices result in the long end-to-end latencies due to the limited PCIe bandwidth; 2) the global memory capacity of an GPU becomes one of the main limitations for the microservice co-location,
because each microservice occupies large global memory space.}

{\color{black}While there is no standard GPU microservice benchmarks, we first develop {\it Camelot suite}, a benchmark suite that includes both real and artifact GPU microservices. The real-system workloads include end-to-end services that cover natural language processing (NLP), deep neural network (DNN) and image processing. We use cutting-edge model such as LSTM~\cite{lstm}, Bert~\cite{bert}, VGG~\cite{vgg}, and DC-GAN~\cite{dcgan} to build the real-system benchmarks, and the benchmarks are programmed with python, C/C++, and CUDA. 
The artifact benchmark is comprised of compute intensive, memory intensive and PCI-e intensive microservices. We can emulate various end-to-end services using the artifact benchmark.
}

Because the load of a user-facing service
 varies (diurnal load pattern~\cite{barroso2009datacenter}) and the contention scenario is only known at 
runtime, an online method is required to manage
the GPU microservices. 
We therefore propose a runtime system named {\bf Camelot} to manage GPU resources online. 
In Camelot, a {\it global memory-based
communication mechanism} enables fast data transfer between 
microservices on the same GPU; 
two {\it contention-aware resource allocation policies} identify the optimal GPU resource allocations that 
minimize the resource usage or maximize the throughput while ensuring the required QoS. 
The allocation decisions are made based on the pipeline effect of microservices and the runtime contention behaviors.
To enable the effective resource allocation, 
we also propose a performance predictor that
precisely predicts the global bandwidth usage, duration, and throughput 
of each microservice under various resource configurations. 
This paper makes three main contributions.




\begin{itemize}
	\item {\bf Comprehensive characterization of GPU microservices.} The characterization
	reveals the challenges in managing GPU microservices. We will open source both the benchmark suite and the runtime system\footnote{The source code is available at github. Currently the link is hidden due to the double-blind review but available by request.}. 

	\item {\bf A global memory-based communication mechanism for GPU microservices.} Adopting the mechanism, 
	the microservices on the same GPU communicate directly without the expensive CPU-GPU data copies. 
	
	\item {\bf A lightweight GPU resource allocation policy.} The policy considers communication overhead, global memory capacity, shared resource contention, and pipeline stall when managing the GPU resources.

\end{itemize}

We implement Camelot and evaluate it on a GPU server with two Nvidia 2080Ti GPUs, and a DGX-2 machine with Nvidia V100 GPUs. 
According to our experimental results, Camelot
 effectively increases the supported peak load by up to 73.9\% and 64.5\% compared with EA and Laius, and 
 reduces the GPU resource usage by 46.5\% compared with equal allocation and 35\% compared with Laius at low load while ensuring the required QoS.


\section{Related work} \label{sec:related}
There have been some efforts on related topics: resource management and scheduling in datacenters, benchmark suites for user-facing services, and microservice architecture.

\textbf{Microservice Architecture}. 
 Yu et al. proposed a microservice benchmark suite DeathStarBench~\cite{gan2019open}, and used it to study the architectural characteristics of microservices. 
 Li et al.~\cite{li2019dataflow} presented a data flow-driven approach to semi-automatically decompose cloud services into microservices. Zhou et al.~\cite{zhou2018poster} identified the gap between existing benchmarks and industrial microservices, and proposed a medium-size microservice benchmark system. There also exist some efforts on the measuring the performance of microservice-based applications~\cite{ueda2016workload,amaral2015performance,hasselbring2016microservices,bao2019performance}. 
 Gribaudo et al.~\cite{gribaudo2017performance} provided a simulation-based approach to explore the impact of microservice-based architectures in terms of performances and dependability, given a desired configuration. However, these researches are for CPU microservices and are not applicable for GPU microservices.

\textbf{Resource scheduling for CPU microservices}. There has been a large amount of prior work on improving the utilization while avoiding QoS violations for CPU microservices. Bao et al.~\cite{bao2019performance} analyzed the performance degradation of microservices from the perspective of service overhead and develops a workflow-based scheduler to minimize end-to-end latency and improves utilization. Based on the characteristics of the workload, HyScale and ATOM~\cite{gias2019atom,kwan2019hyscale} designed resource hybrid controllers that combine horizontal and vertical scaling to dynamically resource division to improve the corresponding time of microservices. Considering the complexity of performance prediction, Seer~\cite{gan2019seer} proposed an online performance prediction system. 


{\color{black}
\textbf{Resource management on GPUs}.
DART~\cite{xiang2019pipelined} employed a pipeline-based scheduling architecture with data parallelism, where heterogeneous CPUs and GPUs are arranged into nodes with different parallelism levels. 
Laius~\cite{zhang2019laius} allocated the computation resource to the co-located applications for maximizing the throughput of batch applications while guaranteeing the required QoS of user-facing services. Baymax~\cite{chen2016baymax} reorders the GPU kernels for ensuring QoS at co-location on time-sharing accelerators. 
However, none of them considers the dependence relationship between microservices as Camelot does. Ignoring the characteristics of the microservice architecture makes results in low resource utilization compared with Camelot. 
}

\section{Representative Microservices} \label{sec:bench}
In this section, we describe {\bf Camelot suite}, a GPU microservice benchmark suite that includes four representative end-to-end user-facing GPU microservices. Besides, we design an artifact benchmark comprised of compute-, memory- and PCIe-intensive microservices for extensive evaluation. 
{\color{black}
We build Camelot suite based on four guidelines:
\begin{itemize}
	\item {\bf Functional integrity } - The benchmarks should reflect the real world requirements, 
		show full functionality, and are deployable on real systems.
	\item {\bf Programming Heterogeneity} - The benchmarks should allow programming language and framework heterogeneity, with each tier developed in the most suitable language, only requiring a well-designed API for microservices to communicate with each other.
	\item {\bf Modularity} - According to Conway's Third Law~\cite{Conway}, the artifact benchmarks should be independent and modularized. 
	This modularity prevents vague boundaries and sets up the ``inter-operate, not integrate'' relationship between the artifact benchmarks.
	\item {\bf Representativeness} - The computational parts of the microservices should come from popular open-source applications and state-of-the-art approaches used in academic and industrial.
\end{itemize}
}


\subsection{Real system GPU Microservices}
According to the above concepts, 
we choose user-facing applications that uses common deep learning techniques and implement them
in microservice architecture. 
Table~\ref{table:bench} lists the end-to-end user-facing applications that cover a wide spectrum of real applications based on GPU microservices.
\begin{table}[h!]
	\caption{End-to-end GPU microservices in Camelot suite}
	\label{table:bench}
	\centering
	\scriptsize
	\begin{tabular}{|c|c|c|c|}
		\hline
		\textbf{Workload} & \textbf{Miroservices} & \textbf{Implementation} & \textbf{Language}\\
		\hline
		\multirow{2}* {Img-to-img~\cite{imagetoimage}} & Face recognition & FR-API~\cite{face} & PYTHON\& \\
		~ & Image enhancement & ~ FSRCNN~\cite{imagetoimage2} & CUDA\\
		\hline
		\multirow{2}* {Img-to-text~\cite{imagetotext1}} & Image feature extraction &  VGG~\cite{vgg} & C++ \& \\
		~ & Image caption & LSTM & CUDA \\
		\hline
		\multirow{2}* {Text-to-imag\cite{texttoimage1}} & Semantic understanding & LSTM~\cite{lstm} & C++\&\\
		~ & Image generation & DC-GAN~\cite{dcgan} & CUDA \\
		\hline
		\multirow{2}* {Text-to-text~\cite{texttotext1}} & Text summarization & BERT~\cite{bert} & PYTHON\&\\
		~ & Text translation & Opennmt~\cite{opennmt} &	 CUDA \\
		\hline
	\end{tabular}
\end{table}

\begin{figure}
	\centering
	\includegraphics[width=.8\columnwidth]{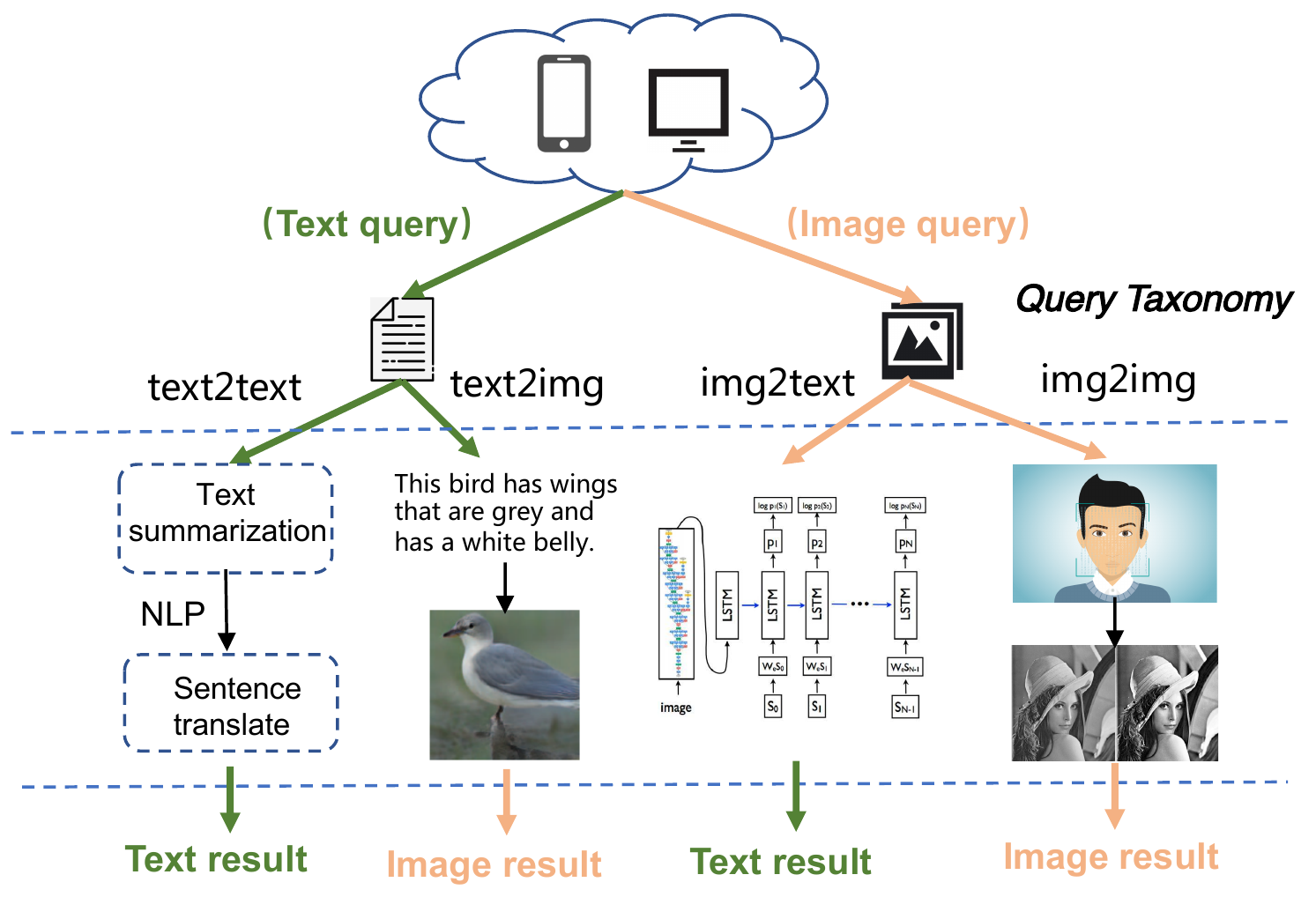}
	\vspace{-2mm}
	\caption{\label{fig:realbench}Tier-level view of the Camelot Suite.}
	\vspace{-4mm}
\end{figure}

Figure~\ref{fig:realbench} illustrates the tiered view of Camelot Suite spanning the query taxonomy it supports, and the end-to-end applications in Camelot Suite. They are widely-used in {\it natural language proessing} (text-to-text), {\it image proessing} (img-to-img), {\it image generation} (text-to-img), and {\it image caption} (img-to-text). Their functionalities are described as follows.

\begin{figure}
	\subfloat[Processing time of the compute-intensive microservice\label{fig:1}]{
		\includegraphics[width=0.47\columnwidth]{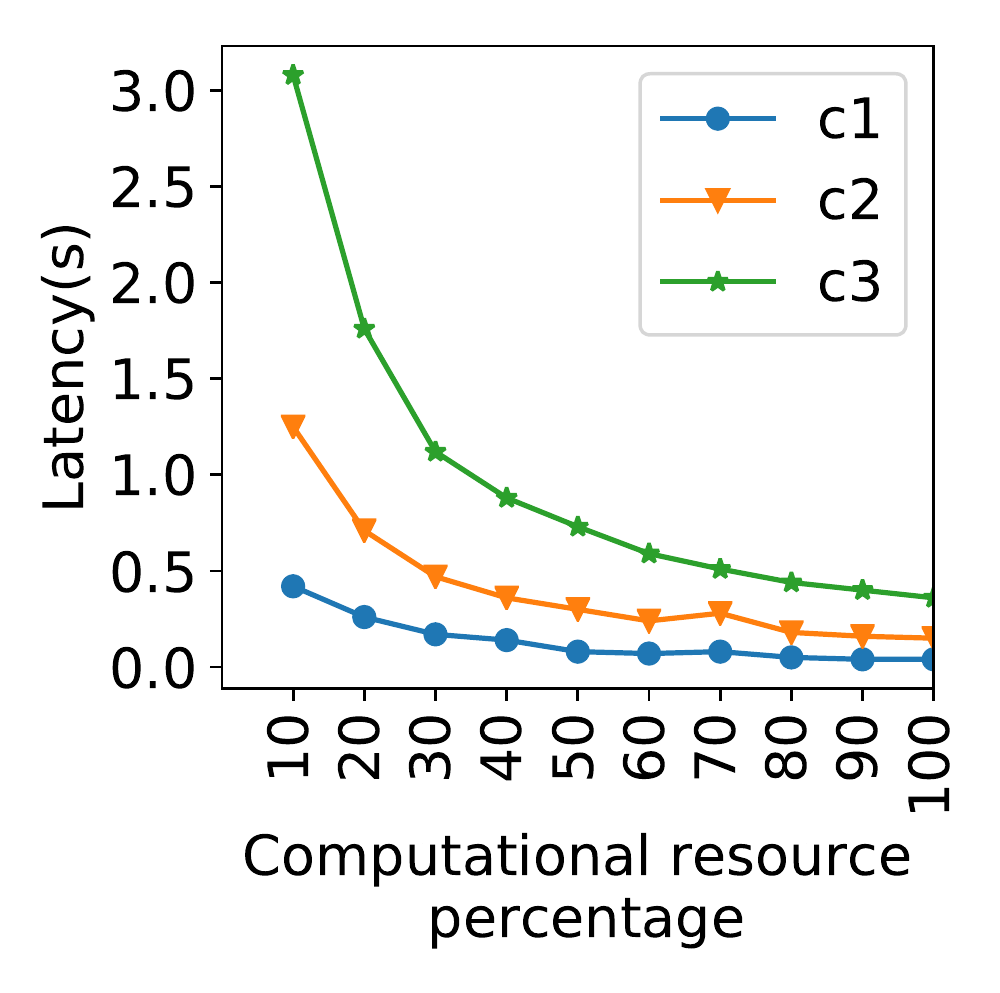}
	}
	\hfill
	\subfloat[Memory bandwidth of the memory-intensive microservice\label{fig:2}]{
		\includegraphics[width=0.47\columnwidth]{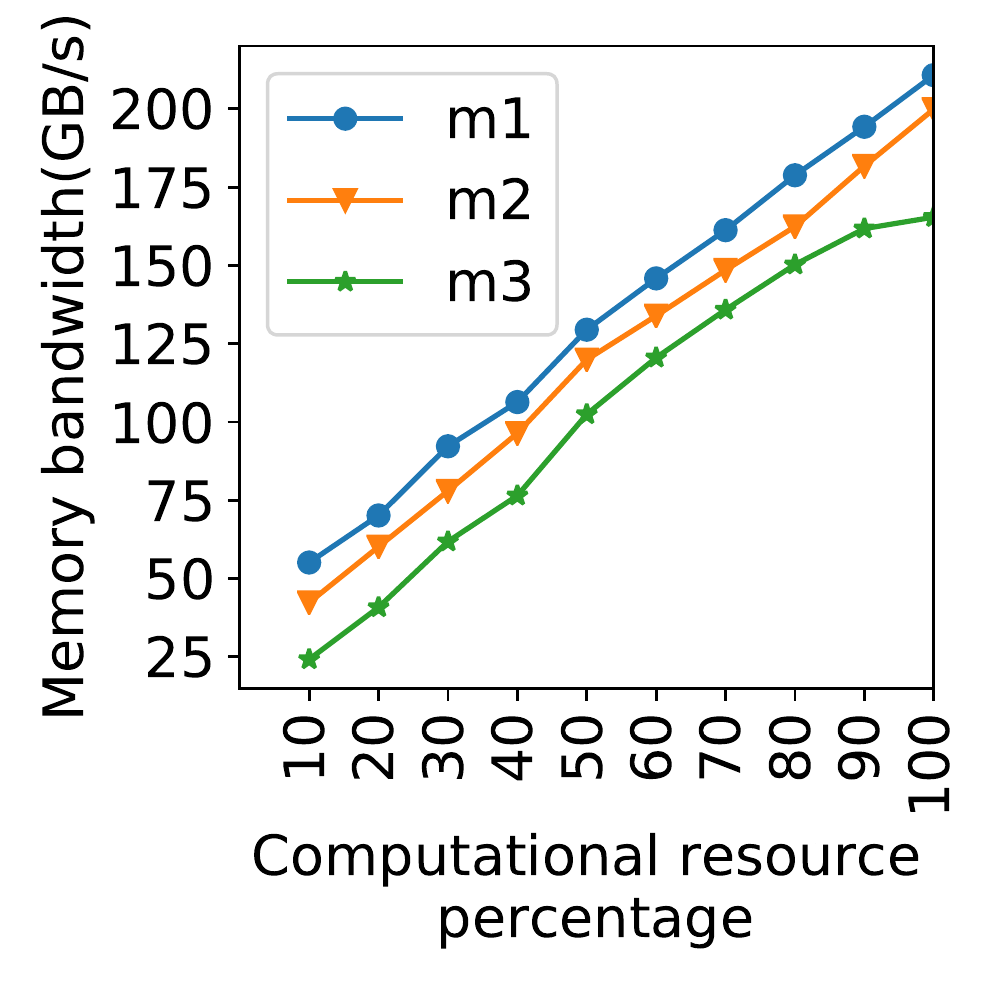}
	}
	\vspace{-2mm}
	\caption{\label{fig:offline_lavamd}Scalability of the artifact benchmarks.}
	\vspace{-3mm}
\end{figure}

{ Natural language processing applications belong to the text-to-text class}~\cite{texttotext1,texttotext2} and consist of two GPU microservices. The first microservice is text summarization task, with Bert. Text summarization is designed to turn text or collections of text into short summaries that contain key information. The second microservice is sentence translation, which aims to translate the text summary output from the first stage into another language.

{ Image processing applications belong to the img-to-img class}~\cite{imagetoimage,imagetoimage1,imagetoimage2}.
The first stage of the application is the face recognition service based on an
open-source project named ``face-recognition''. The second part is the
image enhancement service, which is implemented using the FSRCNN model. This is an image-based microservice where users send requests and upload images to the microservices. Then face recognition service first recognizes the face location information in the image and cut out the tiny faces. Next, the image enhancement service further processes the tiny face to generate a high pixel (64$\times$64) face image.

{ Image generation applications generate new images according to the text, and belong to the text-to-img class}. For example, if
the input to the neural network is ``flowers of pink petals'', the output 
will be an image containing these elements. The task consists of two parts~\cite{texttoimage1,texttoimage2,texttoimage3,texttoimage4,texttoimage5}:
(1) Use natural language processing to understand the description in the
input, with LSTM. (2) Generate a network to output an accurate
image that expresses the text, with deep convolutional generative adversarial network (DC-GAN).

{ Image captioning applications generate a human readable textual 
description for a given image, and belongs to img-to-text class}~\cite{imagetotext1,imagetotext2}. 
The benchmark involves two models: (1)The feature extraction model. 
Given an image, it extracts significant features, which are usually represented by a
vector of fixed length.
VGG is usually used 
for feature extraction. 
(2) The language model. 
For image description, a neural network such as a language
model can predict a sequence of words in a description based on the 
extracted features of the network. 
A common method is to use a cyclic 
neural network, such as Long and Short Term Memory 
Network (LSTM), as a language model. 

%


\subsection{Artifact Benchmarks for Extensive Study}
The artifact benchmarks are ported from three PCI-e intensive, compute-intensive and memory-intensive workloads in Rodinia~\cite{rodinia}. By connecting the artifact benchmarks as needed, we are able to build various end-to-end GPU microservices. 
The arithmetic intensities of the compute intensive microservice and the memory intensive
microservice can be configured accordingly. 
Figure~\ref{fig:offline_lavamd} shows the scalability of the microservices with different compute intensities and memory intensities.
In the figure, $c3$ is configured to be more compute intensive than $c2$ and $c1$, $m1$ is more memory intensive than $m2$ and $m3$. 
The two microservices are sensitive to the resource allocation, thus are suitable to study resource management for general GPU microservices.


\section{Investigating GPU Microservices} \label{sec:background}
We use the real system benchmarks in Camelot suite to investigate the
effectiveness of the current service deployment methods for GPU microservices. 
Specifically, we seek to answer two research questions. 
1) Can the current deployment methods effectively utilize GPU resources?
2) If no, what are the main factors that result in the inefficiency?


\subsection{Inefficient Microservice Pipeline}\label{sec:motivation}
We use two Nvidia RTX 2080Ti GPUs as the experimental platform to perform the investigation. Because our study does not rely on any specific feature of 2080Ti, it applies for other spatial multitasking GPUs. 

\begin{figure}
	\vspace{-2.5mm}
	\subfloat[The supported peak throughput of each microservice if it is assigned a whole GPU.\label{fig:base1}]{
		\includegraphics[width=0.47\columnwidth]{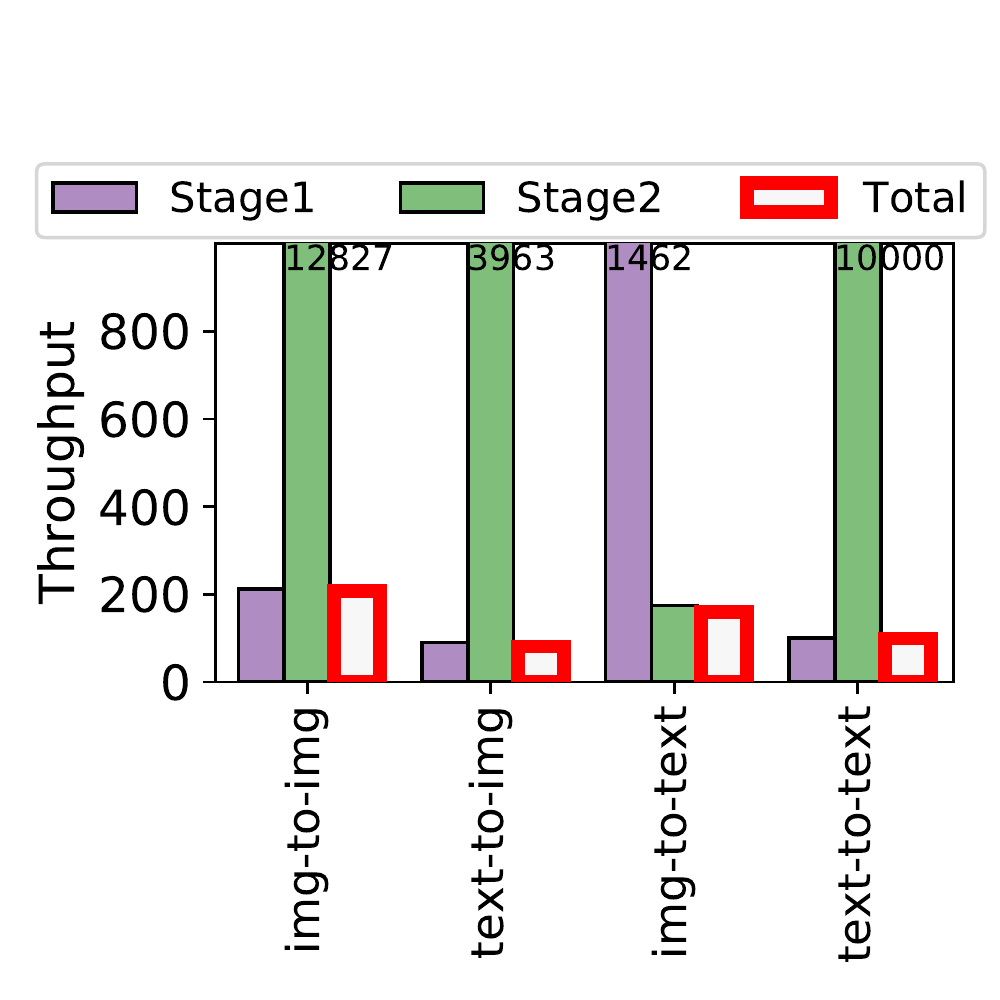}
	}
	\hfill
	\subfloat[The QoS violation of microservices with the balanced deployment policy.\label{fig:base2}]{
		\includegraphics[width=0.47\columnwidth]{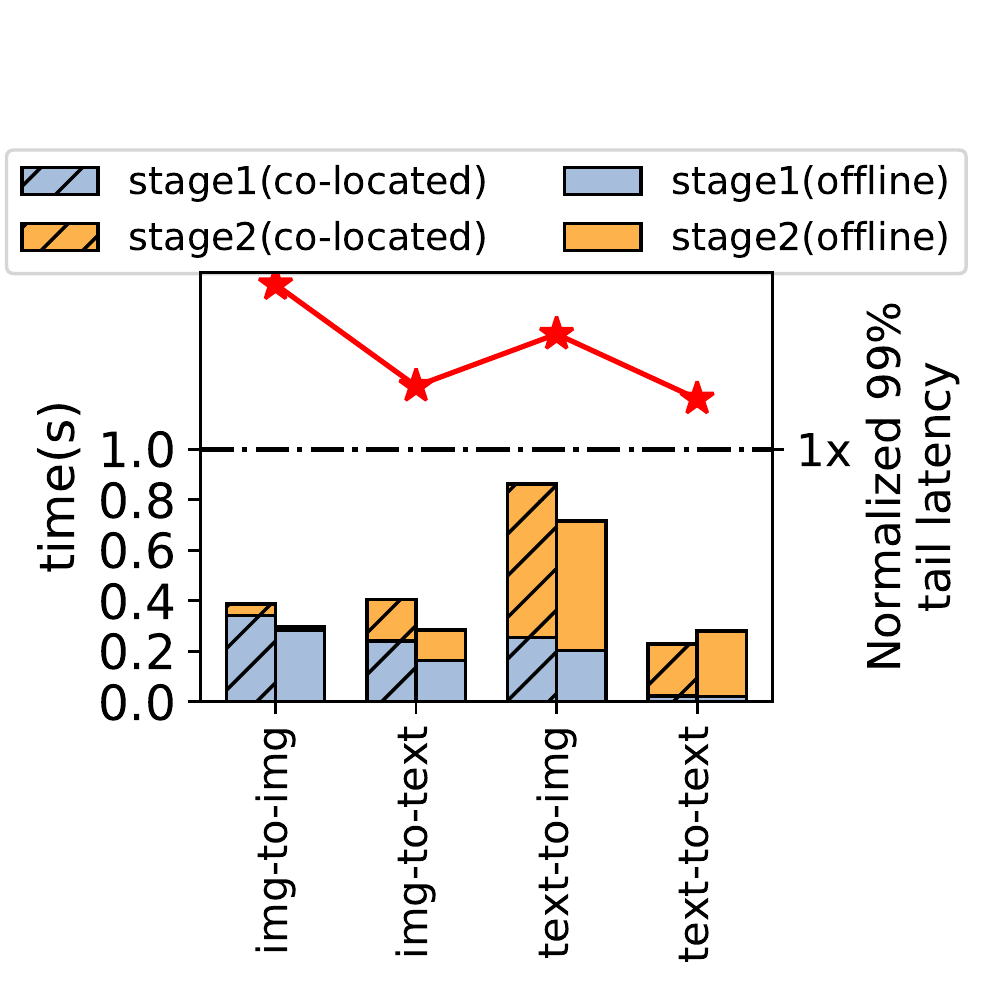}}
	\vspace{-2mm}
	\caption{\label{fig:baseline} Low throughput and QoS violation.}
	\vspace{-3mm}
\end{figure}

{\bf Standalone deployment policy} deploys each microservice on a standalone GPU, and relying on the cross-GPU data copies to perform the communication between the microservices. In this experiment, we gradually increases the load of each benchmark until its 99\%-ile latency achieves the QoS target, and report the peak throughput (i.e., query-per-second, QPS) of the benchmark in Figure~\ref{fig:base1}. 
Let $T_1$ and $T_2$ represent the time spent by a user query on the two microservice stages when the latency of the query reaches the QoS target. 
The bar ``Total'' in the figure shows the peak throughput of the benchmark, ``Stage1'' and ``Stage2'' show the achievable throughputs of the two microservice stages while making sure that their processing time are shorter than $T_1$ and $T_2$ respectively. 

As shown in Figure~\ref{fig:base1}, the peak supported throughput of a benchmark is determined by the microservice stage that has the lowest throughput. For instance, the peak throughputs of {\it image-to-image} and {\it image-to-text} are determined by the first microservice stage and the second microservice stage respectively. Therefore, the standalone deployment policy results in the low peak throughput of GPU microservices due to the inefficient microservice pipeline. This is because it does not consider the differences between the microservices.



\begin{figure}
	\centering	
	\subfloat[Img-to-img]{
		\includegraphics[width=0.11\textwidth]{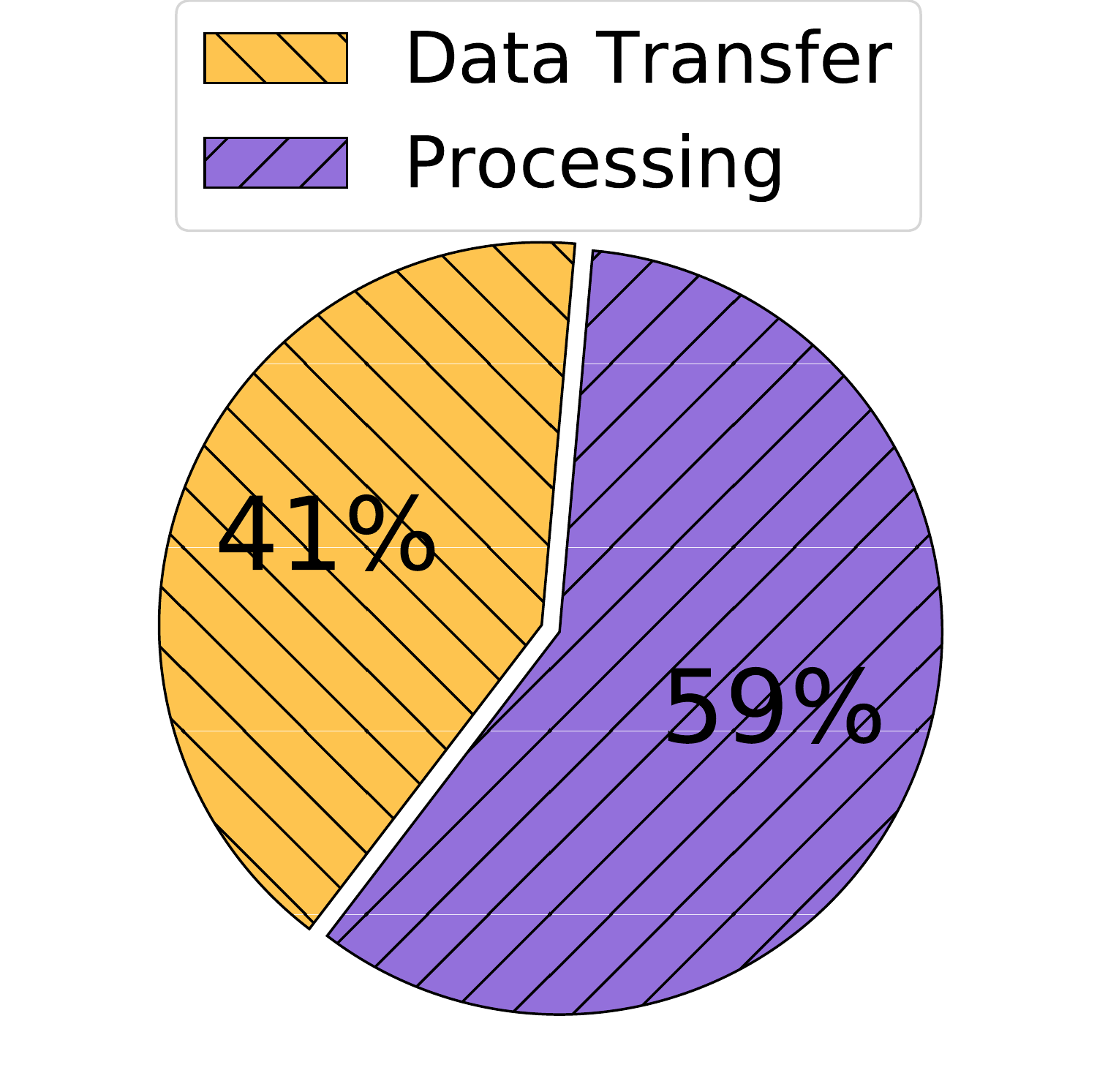}
		\label{fig:1}
	}
	\subfloat[Img-to-text]{
		\includegraphics[width=0.11\textwidth]{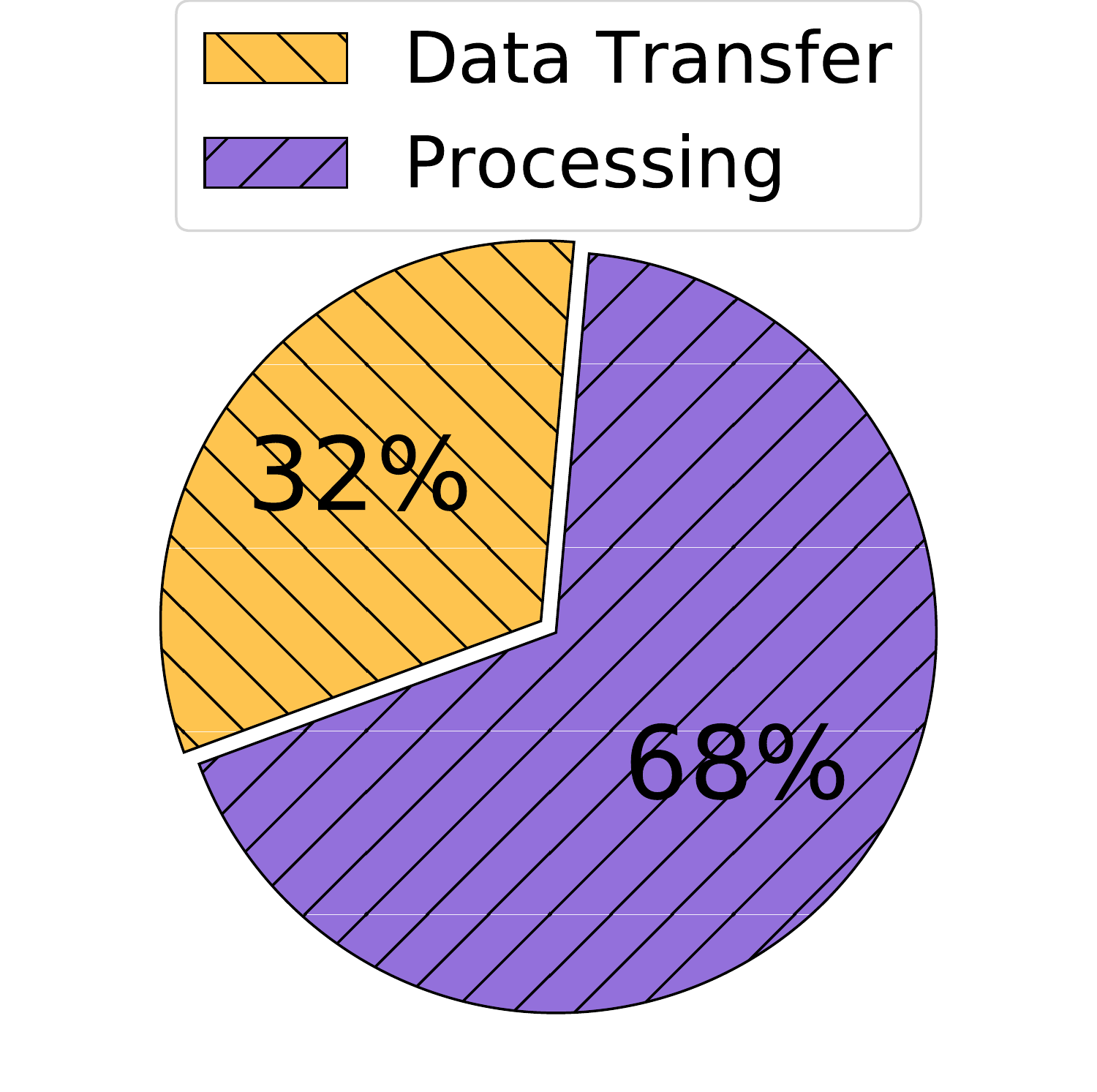}
		\label{fig:2}
	}	
	\subfloat[Text-to-img]{
		\includegraphics[width=0.11\textwidth]{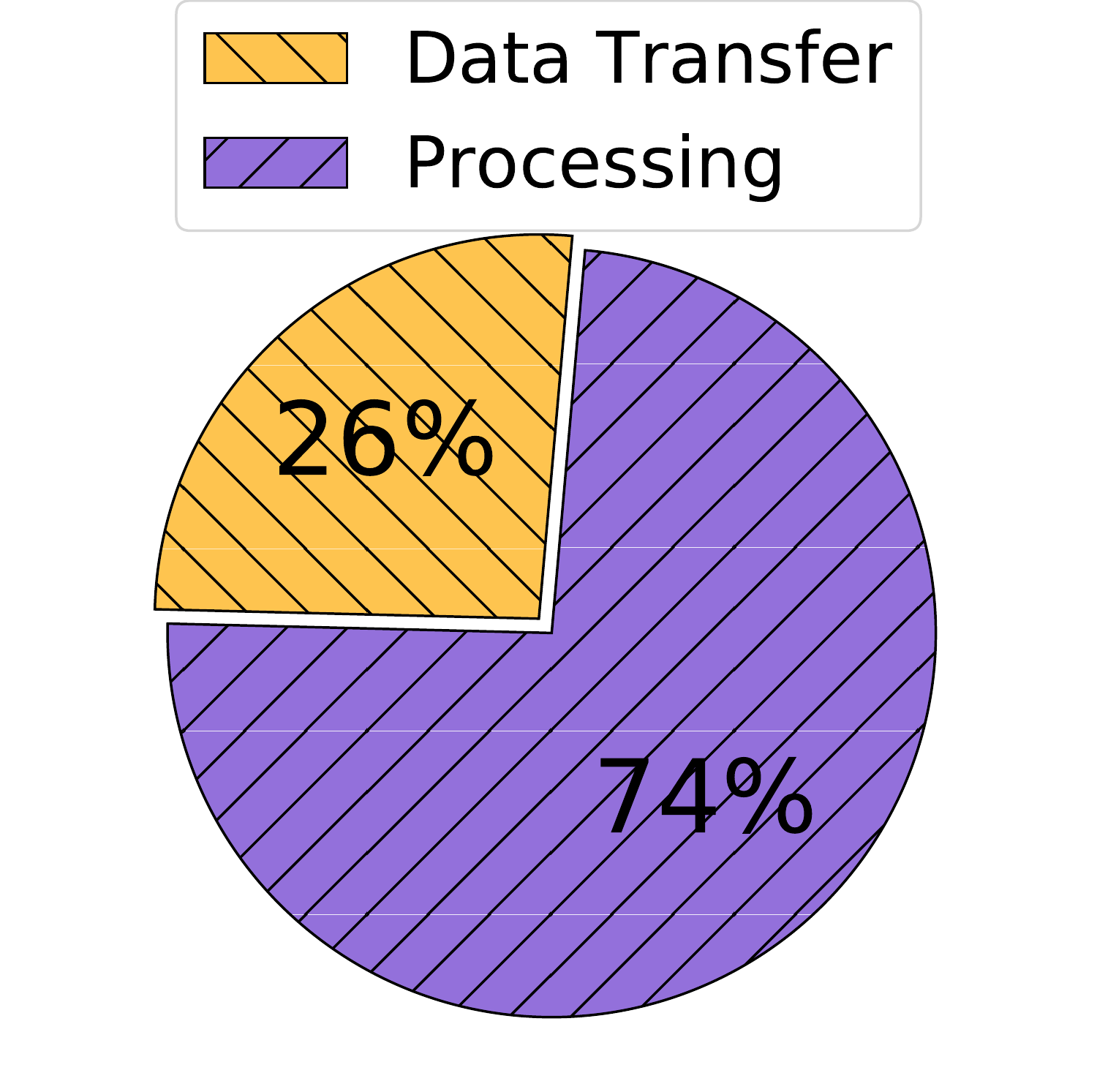}
		\label{fig:3}
	}
	\subfloat[Text-to-text]{
		\includegraphics[width=0.11\textwidth]{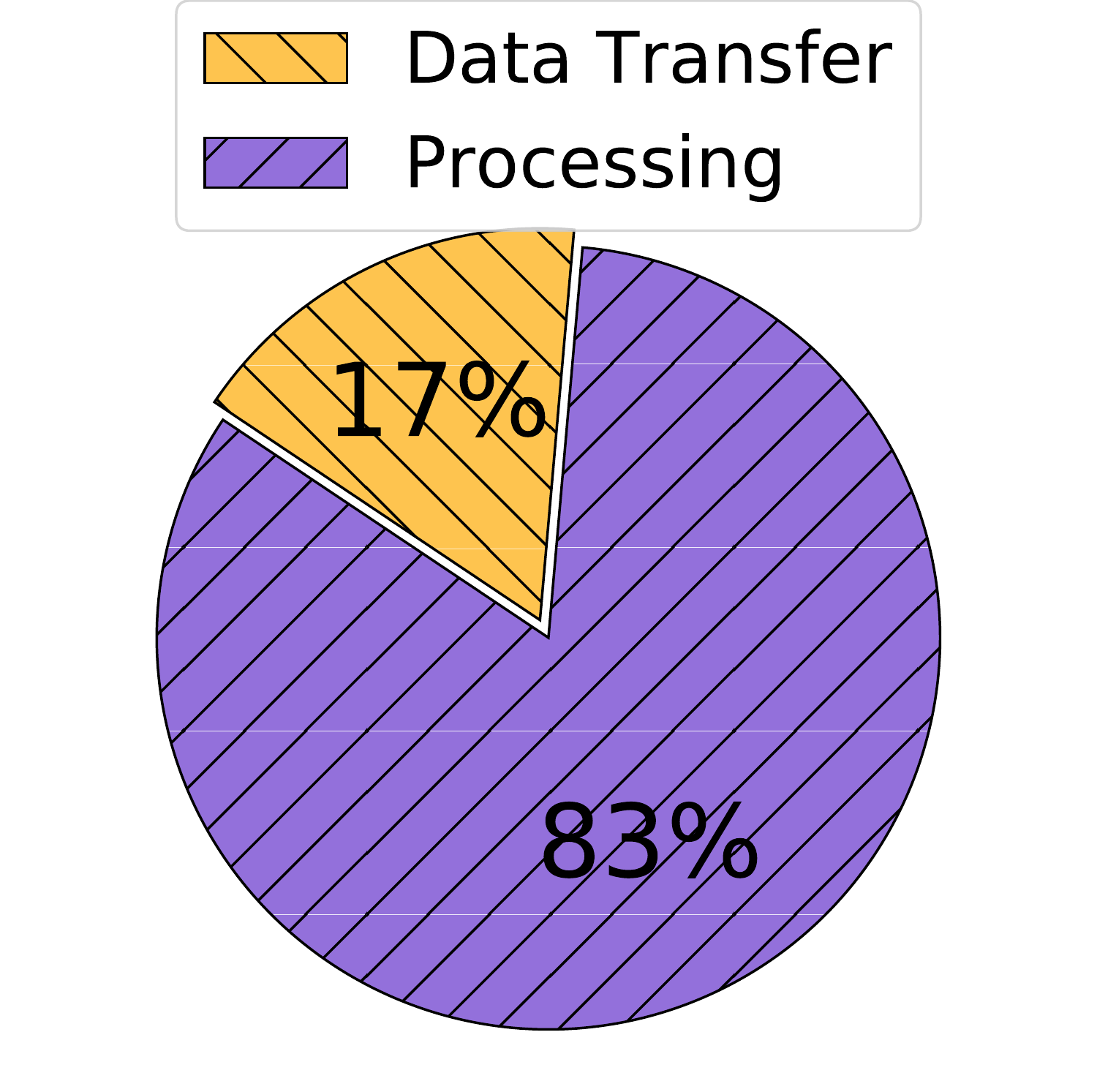}
		\label{fig:4}
	}
	\caption{\label{fig:transfer time} Breaking down the end-to-end latency of a query.}
	\vspace{-4mm}
\end{figure}



{\bf Balanced deployment policy} is designed base on the fact that a user-facing application achieves the highest throughput when the throughputs of its microservice stages are identical. The policy allocates the computational resources (i.e., SMs) to the microservices in a fine-grained manner accordingly. The fine-grained allocation is enabled by the Nvidia Volta MPS technique~\cite{voltamps}. 
To achieve the balanced deployment, for each benchmark, the throughput and processing time of each microservice stage are profiled offline, the SM allocation is carefully tuned so that the throughputs of the two stages are identical, while still ensuring that the aggregated processing time is shorter than the QoS target. For instance, if some SMs of the GPU for Stage2 of the {\it img-to-img} benchmark can be allocated to Stage1, the peak throughput of  {\it img-to-img} can be improved.

{\color{black}Figure~\ref{fig:base2} shows the QoS violation of the benchmarks when the balanced deployment policy is adopted. In the figure, 
the stars represent the normalized 99\%-ile latencies of the benchmarks (corresponding to the right $y$-axis). The bars ``stage1 (offline)'', 
``stage2 (offline)'', ``stage1 (co-located)'', ``stage2 (co-located)'' represent the offline-profiled processing time of the first and the second microservice stages, and the actual processing time of  the first and the second microservice stages respectively (the left $y$-axis). }

We get two observations from Figure~\ref{fig:base2}. As for the first observation, all the benchmarks suffer from QoS violation with the balanced deployment policy. This is mainly because the microservices on the same GPU contend for PCIe bandwidth, global memory bandwidth (Figure~\ref{fig:GPU-interference}), although the SMs are explicitly allocated. The unstable runtime contention behavior results in the long tail latency. As for the second observation, the actual processing time of both the two stages are longer than their offline-profiled processing time due to the shared resource contentions. 
The unbalanced performance degradations due to the contention also result in the inefficient pipleine in consequence.
Our evaluation in Section~\ref{sec:eval-contention} also verifies the necessity to manage global memory bandwidth contention for GPU microservices.

{\it 
The current service deployment policies result in low peak throughputs or QoS violations of GPU microservices due to the inefficient microservice pipeline, without tuning the SM allocation online based on runtime contention behaviors.}

\subsection{Large Communication Overhead}
Besides the inefficient pipeline, the communication overhead between microservices contributes to the long end-to-end latency. As shown in Figure~\ref{fig:GPU-interference}, microservices communicate through the main memory. When a microservice $m_1$ sends the result to the next microservice $m_2$, its data is first transferred from the global memory used by $m_1$ to the main memory, and then transferred back to the global memory used by $m_2$, even if $m_1$ and $m_2$ are on the same GPU. This is because $m_2$ is not allowed to access $m_1$'s data directly.


Figure~\ref{fig:transfer time} shows the breakdown of the end-to-end latencies of the queries in the benchmarks. 
As shown in the figure, the communication time takes a large percentage of the end-to-end latency for all the real applications. The data transfer time (host to device/device to host) takes 32.4\% to 46.9\% of the end-to-end latency. 
If the long communication time is eliminated, we can greatly reduce the end-to-end latency of user queries. In this case, the supported peak load can be further increased, and the required GPU resource decreases to support a low load. 

\begin{figure}
	\centering
	\includegraphics[width=.8\columnwidth]{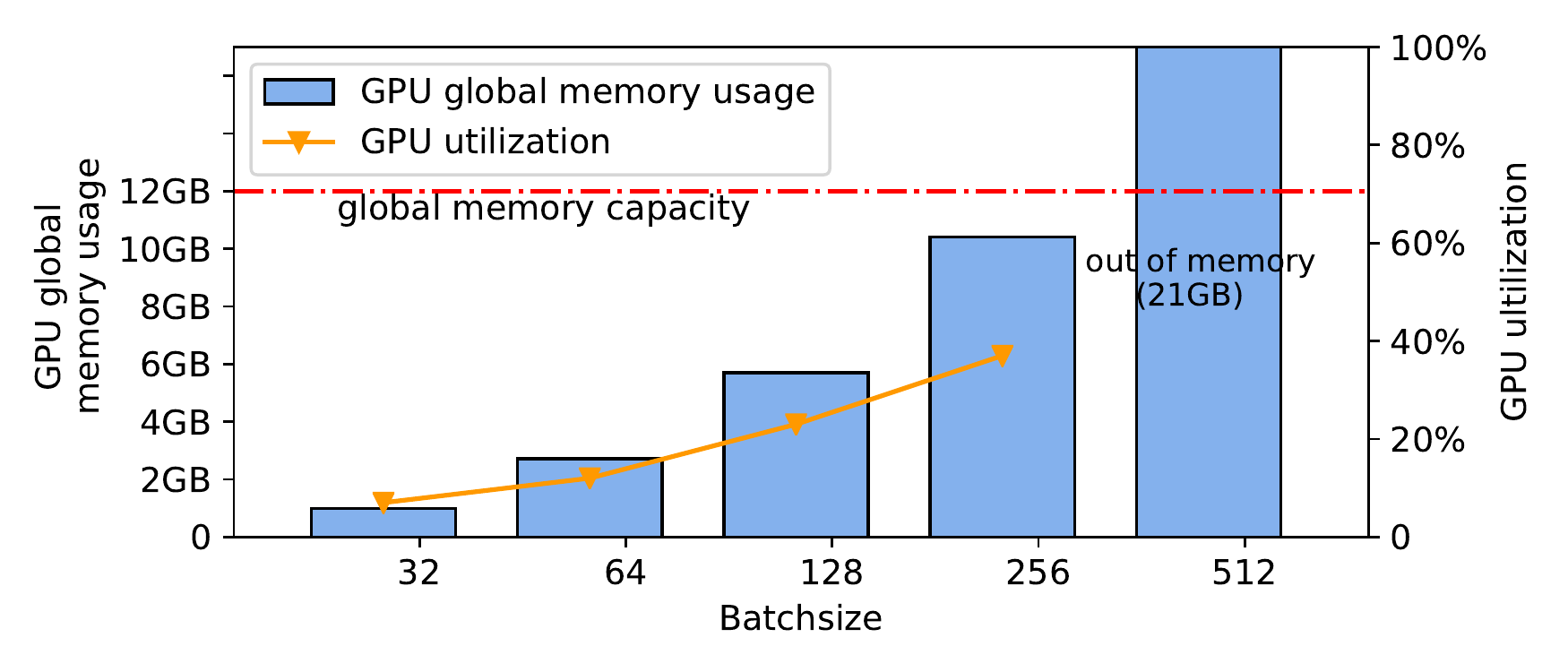}
	\vspace{-3mm}
	\caption{\label{fig:global memory footprint} The global memory usage of the first microservice (FR-API in Table~\ref{table:bench}) in {\it img-to-img} with different batch sizes.}
	\vspace{-3mm}
\end{figure} 

\subsection{Limited Global Memory Space}
While the current machine learning models often use large batch size to improve the throughput, the models are large in capacity. In this scenario, microservices are hard to be co-located on the same GPU due to the limited global memory space. 
As an example, Figure~\ref{fig:global memory footprint} shows the global memory usage and the corresponding GPU utilization when the first microservice of {\it img-to-img} uses different batch sizes. As shown in the figure, the global memory of a GPU is only able to host the microservice with batchsize smaller than 256, while the GPU utilization is lower than 25\%. In this scenario, we are not able to allocate the remaining free computational resource of the GPU to other microservices. 

Unified memory technique that automatic swaps data between the main memory and the global memory can enable the reallocation. However, it incurs heavy data transfer through PCIe bus~\cite{li2015evaluation}. The transfer significantly slows down the communication between microservices (discussed in Section~\ref{sec:CUDAIPC}). 
The limited global memory space of GPUs also contribute to the inefficiency of microservice pipelines. 

Besides the computational resources on each of the GPUs, the resource allocation for improving the pipeline effect of GPU microservices has to consider the global memory space as one of the main constraints.

\section{The Camelot Methodology} \label{sec:overview}
In this section, we propose {\bf Camelot}, a runtime system that maximizes the supported peak load of GPU microservices with limited GPUs and minimizes resource usage at low load while ensuring the QoS requirements.

\subsection{Design Principles of Camelot}\label{sec:challenges}
Based on the investigation in Section~\ref{sec:background}, 
we design Camelot based on three design principles.


{\bf (1) Camelot should minimize the communication overhead between microservices.} 
The CPU-GPU data transfer between microservices results in the long end-to-end latency.
In addition, the PCI-e bandwidth contention between microservices instances also leads to increased 
communication overhead and long latency.

{\bf (2) Camelot should maximize pipeline efficiency while achieving the required QoS online.}
The pipeline efficiency is affected by both the percentage of SM resources allocated to each microservice and 
the runtime contention behaviors, since the microservices on the same GPU contend for the shared resources (e.g., global memory bandwidth).


{\bf (3) Camelot should schedule microservices across multiple GPUs considering the limited global memory space}. 
Since the global memory space is one of the resource bottlenecks for GPU microservices, Camelot should be able to use multiple GPUs to host a end-to-end microservice-based application. Same to the SMs, the GPU memory space is one of the main constraints when scheduling the microservices. 

\subsection{Overview of Camelot}
\begin{figure}
	\centering
	\includegraphics[width=\columnwidth]{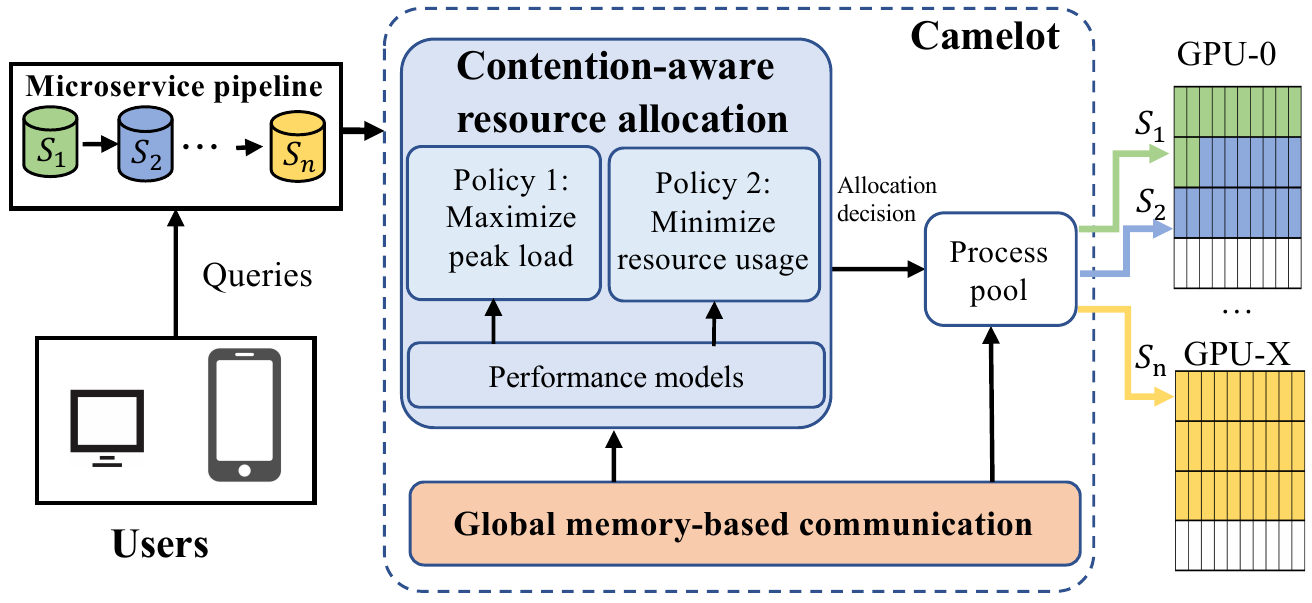}
	\vspace{-5mm}
	\caption{\label{fig:overview}Design overview of Camelot.}
	\vspace{-4mm}
\end{figure}

Figure~\ref{fig:overview} shows the design overview of Camelot. 
As shown in the figure, Camelot adopts a {\it global memory-based communication mechanism} to reduce the communication overhead between microservices on the same GPU. For Camelot, we propose two {\it contention-aware resource allocation policies} that maximize the supported peak load of an end-to-end microservice-based application with limited GPUs and minimize the resource usage at low load respectively, while ensuring the desired 99\%-ile latency target.

The global memory-based communication eliminates the back and forth data transmissions for microservice communication between CPU memory and the global memory of GPU (Section~\ref{sec:CUDAIPC}). It achieves the purpose by only passing the handle of the to be transferred data in the global memory to the receiver. The mechanism resolves the long communication overhead that results in the long end-to-end latency. 

The two resource allocation policies allocate GPU computational resources (i.e., SMs) to the microservices based on the performance of each microservice with various resource configurations, and the runtime contention behaviors (Section~\ref{sec:schedule}). The challenging part here is that Camelot needs to constraint the degradation due to the runtime contention. Otherwise the user-facing service would suffer from QoS violation. By considering both the performance and the contention, the two policies handle the ineffective pipeline effect and the shared resource contention. 

Specifically, when a user query $q$ is submitted, it is processed in the following steps. 1) The query $q$ is pushed into a query wait queue and wait to be issued to the GPU. 2) Once enough queries are received or the first query in the queue tend to suffer from QoS violation, the queries are batched and issued. 3) According to the batch size, Camelot calcualates the GFLOPs (floating point operations) of the batch of queries, and predicts the global memory usage, global memory bandwidth usage, processing time, and throughput (executed requests per second) of each microservice stage under various computational resources. The prediction is done based on an offline-trained performance model. 4) Based on the prediction, Camelot identifies the percentages of the computational resource that should be allocated to each microservice and determines the number of instances for each microservice stage.
5) When co-locating these microservice instances, Camelot considers the reduced communication overhead with the global memory-based communication, the contention on the global memory bandwidth, and the limited global memory space on each GPU. Camelot uses the process pool technique proposed in Laius~\cite{zhang2019laius} to realize the dynamic SM allocation.

\section{Reducing Communication Time}
\label{sec:CUDAIPC}
In this section, we present a global memory-based communication mechanism that enables fast communication between microservices on the same GPU. 

\subsection{Characterizing the Contention on PCIe Bus}
\begin{figure}
	\vspace{-2.5mm}
	\subfloat[The default mechanism\label{fig:1}]{
		\includegraphics[width=0.47\columnwidth]{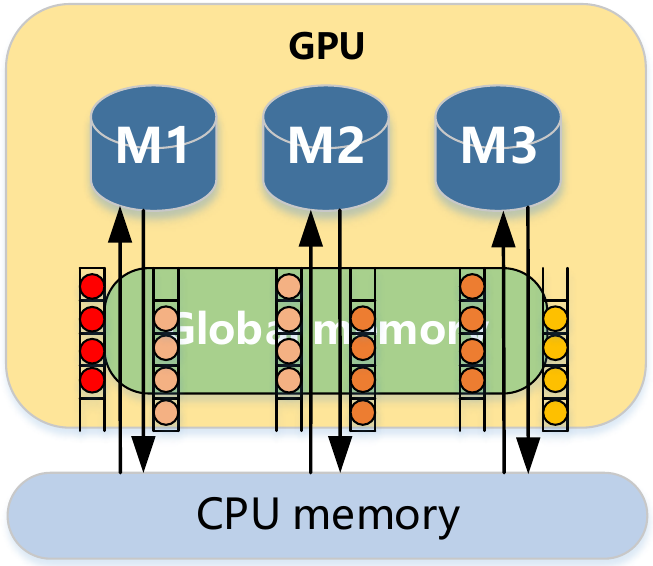}
	}
	\hfill
	\subfloat[The proposed mechanism\label{fig:2}]{
		\includegraphics[width=0.47\columnwidth]{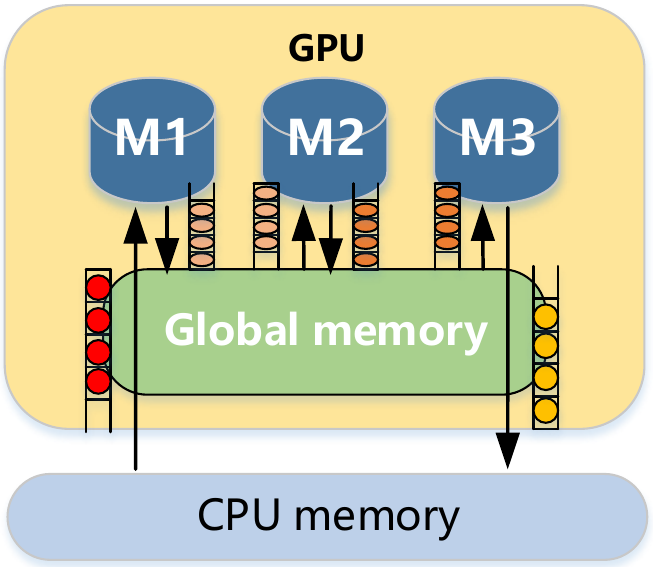}
	}
	\vspace{-2mm}
	\caption{\label{fig:ipc-compare}Comparison of the default and the proposed global memory-based communication mechanisms.}
	\vspace{-4mm}
\end{figure}

Figure~\ref{fig:ipc-compare} compares the traditional main memory-based communication and the proposed global memory-based communication mechanism between microservices. 
During the execution of GPU microservices, since the input of the next stage depends on the output of the previous stage, the results of a microservice stage must be transferred to the next stage. 
As shown in Figure~\ref{fig:ipc-compare}(a), adjacent microservices in the pipeline (e.g., $M1$ and $M2$, $M2$ and $M3$) communicate with each other by copying data back and forth between GPU global memory and the CPU memory. The default communication mechanism results in the long communication latency and the low data transfer bandwidth, especially when multiple microservices co-run on the same GPU.

\begin{figure}
	\centering
	\includegraphics[width=0.7\columnwidth]{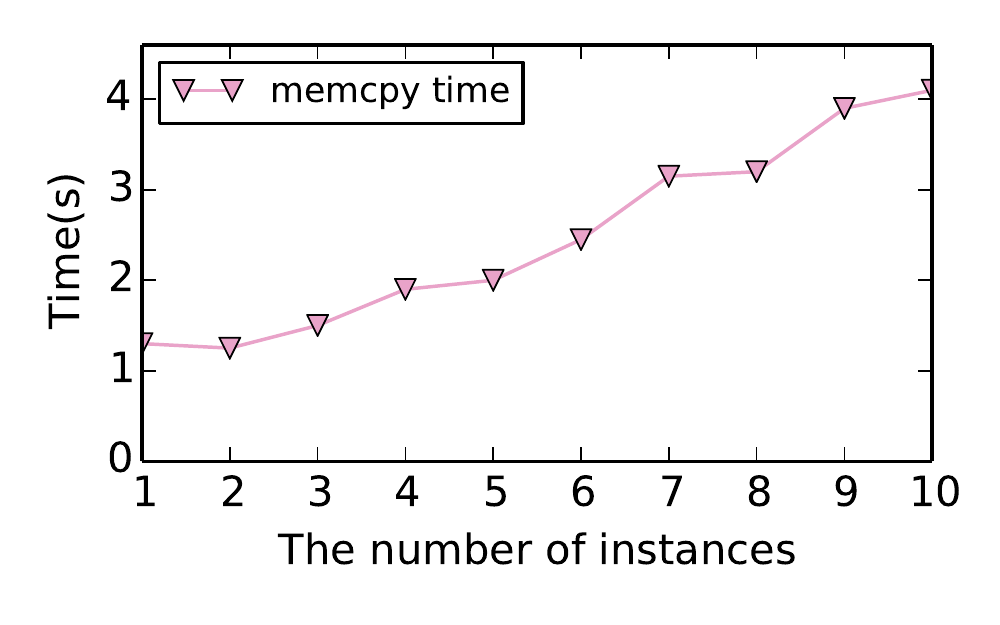}
	\vspace{-4mm}
	\caption{\label{fig:analog bench}The kernel processing time and the PCIe transfer time for PCIe-intensive microservice.}
	\vspace{-4mm}
\end{figure}

To show the impact of the default communication mechanism, we perform an experiment that runs multiple instances of a PCIe-intensive microservice $M$ concurrently on a GPU. {\color{black} 
The functionality of $M$ is copying 5GB data from the main memory to the global memory.} In the experiment, each instance of $M$ is allocated only 10\% of the computational resource to eliminate the impact of the contention on the SMs. 
Figure~\ref{fig:analog bench} shows the data transfer time over PCIe bus of an instance of $M$. In the figure, the $x$-axis shows the number of the instances of $M$ on the GPU.

As shown in Figure~\ref{fig:analog bench}, the data transfer time increases when more than three instances are co-located. The increased data transfer time is due to the contention on the PCIe bandwidth. While the theoretical peak bandwidth of 16x PCI-e 3.0 bus used in our platform is 15,800MB/s
and the effective bandwidth is 12,160MB/s~\cite{goldhammer2008understanding}, and a single memcpy task uses PCIe bandwidth of 3,150MB/s according to our measurement. If the memcpy task transfers data from pinned memory, a single such memcpy task is able to consume all the PCIe bandwidth.


In this scenario, if a GPU hosts more than $\lfloor \frac{12160}{3150} \rfloor =3$ PCIe-intensive microservice instances, the microservices contend for the limited PCIe bandwidth and suffer from the long communication time. 
The long communication time results in the long end-to-end latency of user queries. 

\subsection{Global Memory-Based Communication}
Observed from Figure~\ref{fig:ipc-compare}(a), the data that should be passed from $M1$ to $M2$ is already in the global memory space of $M1$, although the data is not accessible for $M2$. If $M1$ is able to share the data with $M2$, the expensive memcpy (from device to host, and from host to device) can be eliminated. We design a {\it global memory-based communication mechanism} as shown in Figure~\ref{fig:ipc-compare}(b) to achieve this purpose. In more detail, adopting the global memory-based mechanism, the result of a microservice $M$ is temporarily stored in the global memory. Another microservice is able to access the data from the global memory directly without copying data back and forth between the global memory and the main memory.


 \begin{figure}
	\centering
	\includegraphics[width=.8\columnwidth]{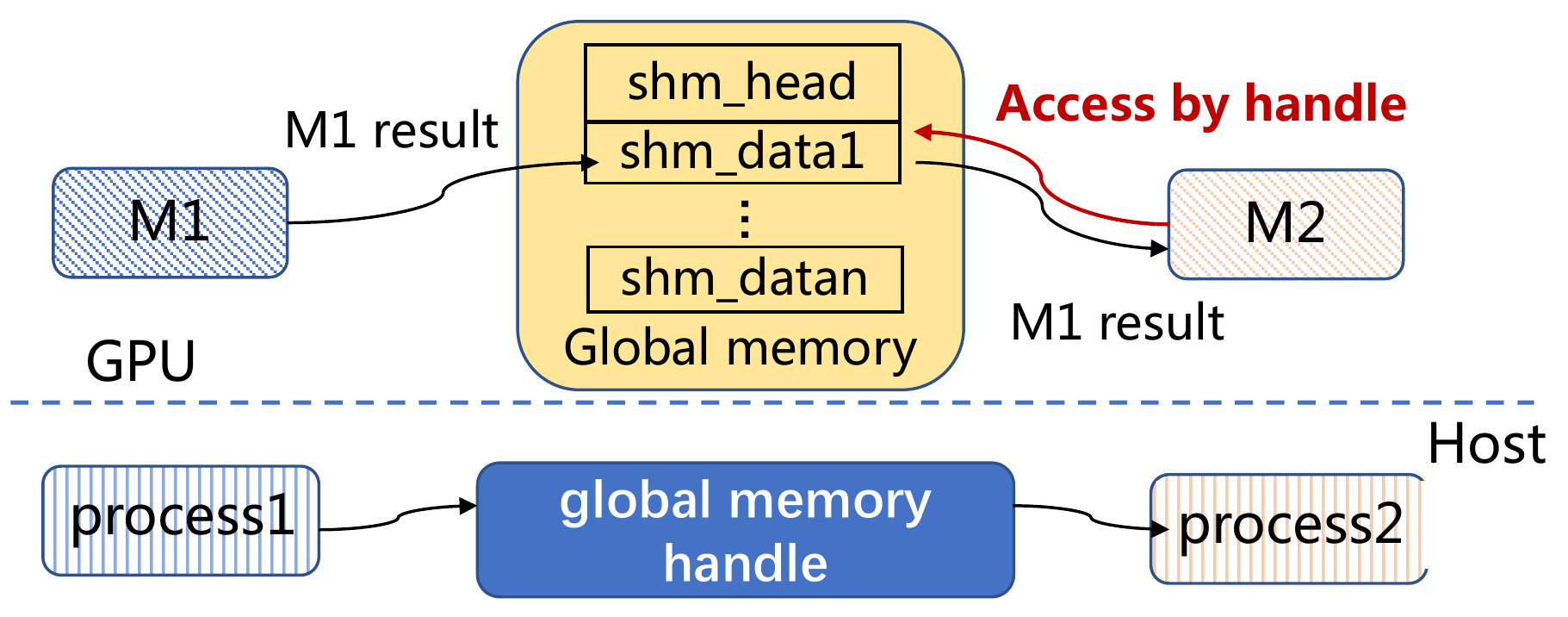}
	\vspace{-3mm}
	\caption{\label{fig:gpu-communication} The global memory-based communication.}
	\vspace{-2mm}
\end{figure} 


Figure~\ref{fig:gpu-communication} illustrates the design of global memory-based communication mechanism. As shown in the figure, 
when a microservice $M1$ needs to pass its result to microservice $M2$ on the same GPU, its process on the host passes a global memory handle (8 bytes) to the process of $M2$ on the CPU side. Once $M2$ gets the data handle, it is able to directly access the data from the global memory. 
We implement the mechanism using the CUDA IPC (inter-process communication) technique provided by Nvidia. The sender process gets the IPC handle for a given global memory pointer using cudaIpcGetMemHandle(), passes it to the receiver process using standard IPC mechanisms on the host side, and the receiver process uses cudaIpcOpenMemHandle() to retrieve the device pointer from the IPC handle.

\begin{figure}
	\centering
	\includegraphics[width=.7\columnwidth]{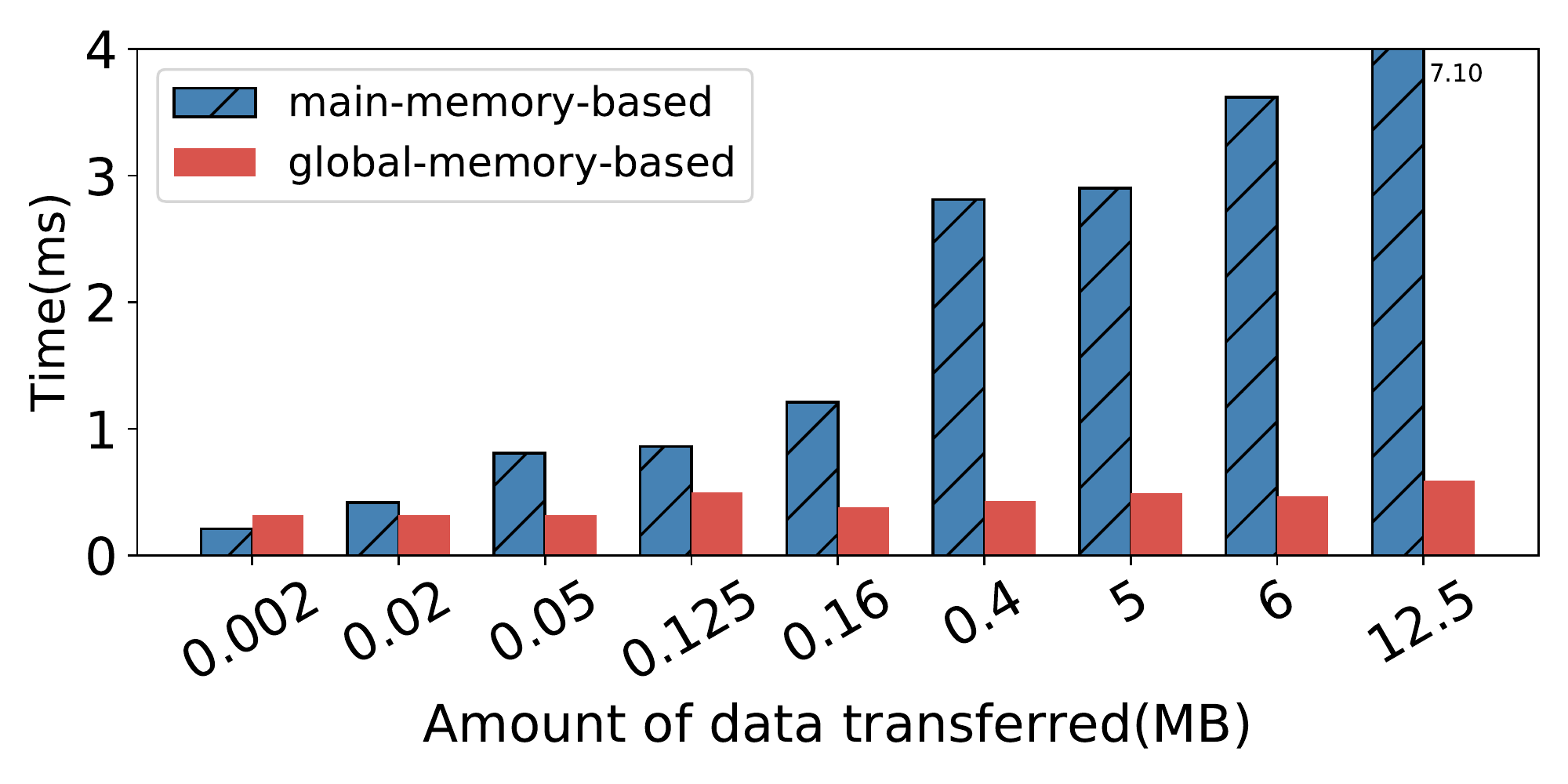}
	\vspace{-3mm}
	\caption{\label{fig:ipc} Communicating with the main memory-based and global memory-based mechanisms.}
	\vspace{-6mm}
\end{figure}

Figure~\ref{fig:ipc} shows the communication time between two microservices on the same GPU using the default and the global memory-based mechanisms.
In the figure, the two microservices do not contend for the PCIe bandwidth. Observed from this figure, the global memory-based mechanism greatly reduces the communication time when the to be passed data is larger than 0.02MB. The larger the to be transferred data, the larger the performance gain is achieved with the global memory-based mechanism. In addition, if the to be transferred data between two microservices are small (e.g., only 2 bytes), the traditional memory-based mechanism shows shorter time. This is mainly because CUDA IPC incurs slight fix overhead when probing, transferring, and decoding the IPC handle in the global memory-based communication mechanism.

Besides reducing the communication time, the mechanism also reduces the global memory space usage of the microservices. 
With the traditional mechanism, $M1$ and $M2$ save two copies of the transferred data. On the contrary, with the global memory-based mechanism, only $M1$  saves a single copy of the transferred data. $M1$ and $M2$ also save a IPC handle of 8 bytes respectively. While the transferred data between microservices are often larger than 8 bytes, the global memory-based mechanism does not consume extra global memory space. Instead, it reduces the global memory usage.

It is worth noting that the microservices on different GPUs are not able to communicate through the global memory-based mechanism. 
Therefore, the microservices that require heavy communication should be placed on the same GPU. 

\section{Allocating GPU Resources}
\label{sec:schedule}
In this section, we present two contention-aware resource
allocation policies for GPU microservices. 
The first policy maximizes the supported peak load of GPU microservices with limited GPUs 
while avoiding QoS violation. 
The second policy minimizes GPU resource usage while ensuring the QoS, in case that the load of a service is low. 

\subsection{Low Overhead Performance Prediction}
\label{sec:prediction}

Camelot predicts the  {\it processing duration}, the {\it global memory bandwidth usage}, and the {\it throughput} of each microservice to support the two resource allocation policies. The throughput represents the number of queries that can be processed per second at a microservice. For each microservice, we train it a performance model that predicts its processing duration, global memory bandwidth usage, and throughput.


The model for a microservice takes its {\it input batchsize} and {\it percentage of computational resources} as the input features, as they seriously affect the microservice's performance. 
The input batchsize reflects the workload of a query, and the percentage of computational resource reflects the computational ability used to process the query. And these features can be collected by profiling tools such as Nsight Compute provided by Nvidia~\cite{nsight}. 
To collect training samples for a microservice, we submit queries with different batch sizes, execute them with different computational resource quotas and collect the corresponding duration. During the profiling, queries are executed in solo-run mode to avoid interference due to shared resource contention. 

Since the QoS target of a user query is hundreds of milliseconds to support smooth user interaction~\cite{tail}, it is crucial to choose the modeling technique that shows both high accuracy and low complexity. We evaluate a spectrum of broadly used low latency algorithms for the microservice performance prediction: Linear regression (LR)~\cite{lr}, Decision Tree (DT)~\cite{dt}, and Random forest (RF)~\cite{rf}. 
\begin{figure}
	\centering
	\includegraphics[width=.6\columnwidth,height=3cm]{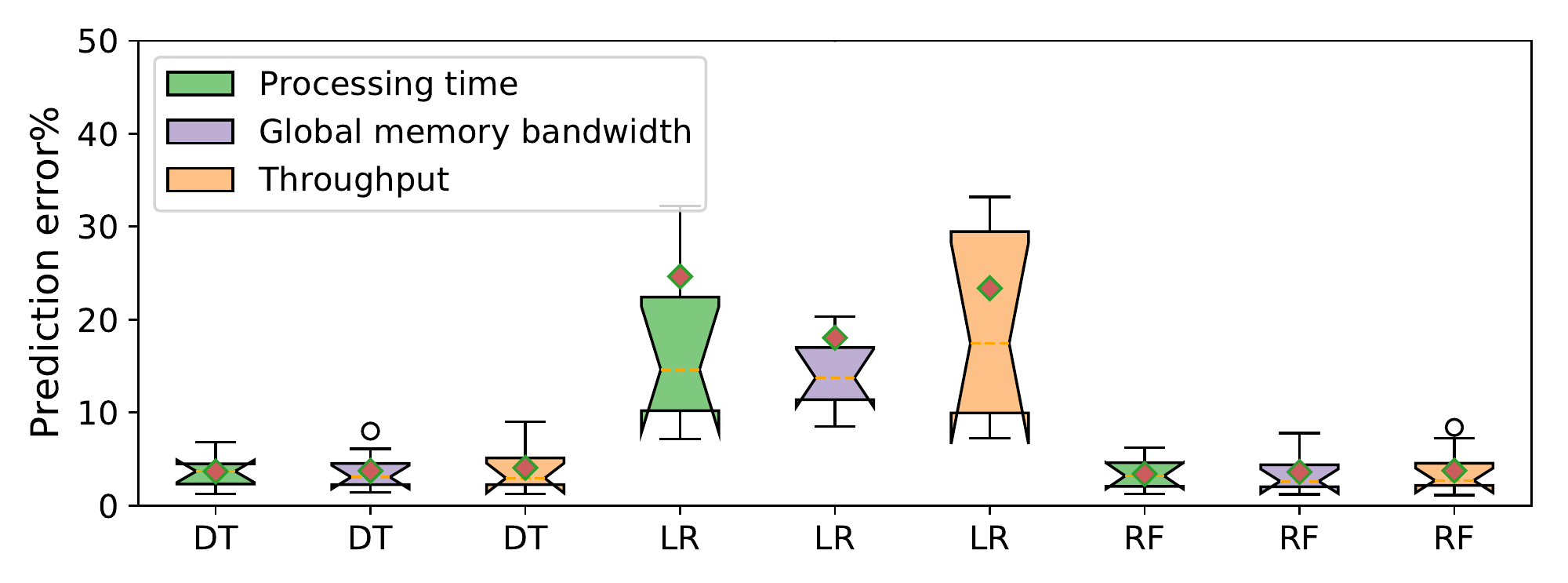}
	\vspace{-4mm}
	\caption{\label{fig:prediction} Errors of predicting duration, global
	memory bandwidth and throughput with DT, LR, and RF.}
		\vspace{-4mm}
\end{figure} 

To evaluate the accuracy of the three modeling techniques, we use 70\% of the collected samples to train the models and use the rest for testing. Figure~\ref{fig:prediction} present the prediction errors of the duration, global memory bandwidth
usage, and throughput of each microservice in Camelot suite with LR, DT, and RF. 
In general, DR and RF show high accuracy for predicting the microservice performance. 
Besides accuracy, we also measure the execution time of different prediction models. The time
of predicting with {\it DT} is shorter than 1 {\it ms},
while the the RF model runs higher than 5 {\it ms}.
We therefore choose DT as the modeling technique to train the performance models. 
Besides, 
Camelot also predicts the FLOPs (floating point operations) and the required
global memory space of the microserives with different workloads. 
LR is able to precisely capture such linear relationship.

We do not use black box methods, such as Reinforcement Learning~\cite{kaelbling1996reinforcement} or Bayesian optimization~\cite{snoek2012practical}, to predict the performance of microservices online, because in-production GPUs lacks the ability to obtain runtime statistics online with low overhead. 
 In a datacenter, it is acceptable to profile a service and build a new model before running it permanently. Similar to prior work on datacenters~\cite{zhang2019laius}, the profiling is done offline so it does not incur runtime overhead.

\subsection{Case 1: Maximizing Peak Load}
\label{sec:case1}

The peak load of an end-to-end service is determined by the smallest peak load of its microservices. 
Therefore, the design principle here is maximizing the smallest throughput of the microservices in an end-to-end service, while still ensuring the end-to-end latency shorter than the QoS target. Camelot tunes the number of microservice instances for each microservice stage, and the SM resource quota for each microservice instance. Other resources (such as global memory bandwidth)  cannot be explicitly allocated. 

We formalize the above problem to be a single-objective optimization problem, where the objective function is maximizing the smallest throughput of the microservices, and the constraints are {\it global memory capacity}, {\it global memory bandwidth}, {\it computational resources on the GPUs} and {\it the QoS target of microservices}. 
In addition, the number of instances for each microservice stage and the resource quotas allocated to each process can be derived from an optimization problem related to its feasible solutions. 


The constraints in the optimization problem are as follows. 
First, to avoid global memory bandwidth 
contention, the accumulated global memory bandwidth required by 
all the microservices on a GPU should be less than the available 
global memory bandwidth of the GPU. Second, the accumulated computational resource quotas 
allocated to concurrent instances should not exceed the overall available computational resources. 
Third, the number of microservice instances on a GPU should not exceed 48 (
Volta MPS allows at most 48 client-server connections per-device).  Fourth, the total time required for 
the total user-facing application should be smaller than the QoS target. 
\begin{table}
\caption{The variables used in the optimization problem}
\label{table:Variable}
\centering
\scriptsize
    \begin{tabular}{|c|c|c|}
      \hline
      \textbf{Variable}& \textbf{Varible description} & Provided by\\
      \hline
      $A_i$ & the $i$th part of Microservice $A$ & Benchmarks\\
      \hline
      $p_i$& the computational resource quotas & \\& allocated to the $i$th microservice & Section~\ref{sec:case1}\\
      \hline
      $s$ & the batchsize of Microservice $A_i$ & scheduler \\
      \hline
      $C$& the total number of GPUs & Section~\ref{sec:case2}\\
      \hline
      $BW$& the available global memory bandwidth & Nvprof\\
      \hline
      $I$& the maximal client CUDA contexts & \\&supported by Volta MPS server per-device & Volta MPS\\
      \hline
      $R$ & the overall computational resources respectively & Nvprof\\
      \hline
      $N_i$ & the number of the $i$-th microservice's processes & scheduler\\
      \hline
      $f(p_i)$ & the predicted throughput of $A_i$ & Section~\ref{sec:prediction}\\
      \hline
      $g(p_{i})$ & the predicted global memory bandwidth & \\& usage of $A_i$ & Section~\ref{sec:prediction}\\
      \hline
      $M(i,s)$ & the global memory footprint of $A_i$ & \\& with batch size $s$ & Section~\ref{sec:prediction}\\
      \hline
      $C(i,x)$ & the amount of calculations of $A_i$ & \\& with batchsize $s$ & Section~\ref{sec:prediction}\\
      \hline
      $G$ & the GFLOPS of the used GPU & Nsight compute\\
      \hline
      $F$ & the global memory capacity of the used GPU & Nvprof\\
      \hline
    \end{tabular}
    \vspace{-4mm}
  \end{table}
  
Assume a user-facing GPU application $Q$ has $n$ microservice stages. Equation~\ref{eq:formalize1} shows the object and the constraints in the optimization problem. 
Table~\ref{table:Variable} lists the variables used to maximize the supported peak load by solving the optimization problem in Equation~\ref{eq:formalize1}.
\begin{equation}
\scriptsize
\begin{aligned}
\label{eq:formalize1}
\text{Object:} & \qquad MAX(\min\nolimits_{i=1}^n N_i\times f(p_{i})), \\
\text{Constraint-1:} & \qquad \sum\nolimits_{i=1}^n N_i \times p_i \leq C*R, \quad 0\leq p_i\leq R\\
\text{Constraint-2:} & \qquad \sum\nolimits_{i=1}^n N_i \leq C\times I, \quad 0\leq N_i\leq I  \\
\text{Constraint-3:} & \qquad \sum\nolimits_{i=1}^n N_i \times b(p_{i})\leq BW \\
\text{Constraint-4:} & \qquad \sum\nolimits_{i=1}^n N_i \times M(i,s))\leq F \\
\text{Constraint-5:} & \qquad \sum\nolimits_{i=1}^n g(p_{i}) \leq QoS \\
\end{aligned}
\end{equation}


\subsection{Case 2: Minimizing Resource Usage}
\label{sec:case2}
In this policy, Camelot first minimizes the number of GPUs required to support the low load, and then minimizes the resource usage in each of the GPUs.
This design choice is able to reduce the search space for resolving the optimization problem described later. 

To determine the minimum number of GPUs required, Camelot already predicts the number of floating point operations and the global 
memory footprint of microservices with different loads (C(i,s) and M(i,s) in Table~\ref{table:Variable}). 
Based on the designed GFLOPS (Giga floating-point operations per second) and the global memory capacity of a GPU, 
Equation~\ref{eq:formalize} calculates the minimum number of GPUs $y$ required. 
In the equation, $G$ and $F$ represent the GFLOPS and the global memory capacity of the used GPU respectively.
Observed from the equation, the minimum number of GPUs required is calculated under constraints of both the computational ability and the global memory space.
\begin{equation}
\label{eq:formalize}
\scriptsize
y= MAX(\frac{\sum\nolimits_{i=1}^n C(i,s)}{G},\frac{\sum\nolimits_{i=1}^n M(i,s)}{F} )
\end{equation}

Equation~\ref{eq:formalize2} shows the object and the constraints 
that further reduces the resource usage in the $y$ GPUs. {\color{black}When choosing the batch size to run microservices, Camelot considers the global memory footprint of different batch sizes(M(i,s)). When global memory resources are scarce, excessive batch size will put pressure on the global memory space. Therefore, batch size should also be considered as a variable when determining resource allocation in the next stage.}
\begin{equation}
\scriptsize
\begin{aligned}
\label{eq:formalize2}
\text{Object:} &\qquad MIN(\sum\nolimits_{i=1}^n N_i \times p_i), \quad 0\leq p_i\leq R \\
\text{Constraint-1:} & \qquad \sum\nolimits_{i=1}^n N_i \leq I, \quad 0\leq N_i\leq I  \\
\text{Constraint-2:} & \qquad \sum\nolimits_{i=1}^n N_i \times b(p_{i})\leq BW \\
\text{Constraint-3:} & \qquad \sum\nolimits_{i=1}^n N_i \times M(i,s))\leq F \\
\text{Constraint-4:} & \qquad \sum\nolimits_{i=1}^n g(p_{i}) \leq QoS \\
\end{aligned}
\end{equation}

\begin{figure}
	\centering
	\includegraphics[width=\columnwidth]{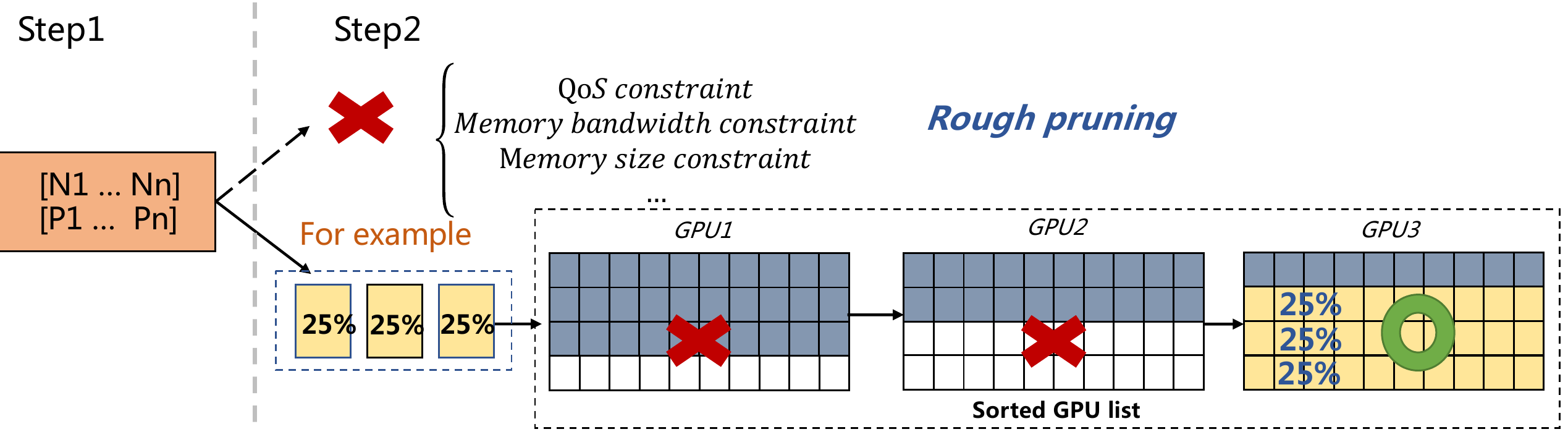}
	\vspace{-6mm}
	\caption{\label{fig:deployment}The deployment scheme of Camelot.}
	\vspace{-6mm}
\end{figure}

{\color{black} By solving the two optimization problems, Camelot finds out the resource quota for each microservice stage, and the number of instances for each microservice stage. We currently adopt Simulated Annealing algorithm~\cite{van1987simulated} to resolve the optimization problems.

In more detail, we have a vector of length 2N called $V$: [n1, n2, .., p1, p2 ...pn], where N is the number of microservice stages. 
For microservice stage-i, we will deploy ni instances and will allocate pi percentage computing resources for each instance. The amount of computing resources of the entire GPU is 100\%. 
Similar to the traditional simulated annealing algorithm, Camelot iterates continuously to search for an optimal result for $V$. 
In each iteration, the current state ($V$) randomly moves in one direction and get a new state candidate ($V'$). 
Camelot will check if the new state $V'$  meets constraints such as memory bandwidth (as shown in formula~\ref{eq:formalize2}). 
If the new state is valid, Camelot calculates the throughput of the new state and compares it with the global optimal throughput. If the new state's throughput is higher, Camelot updates the global optimal throughput. 
If not, Camelot still has the possibility (Acceptance Probability) to update the global optimal throughput as the new state ($V'$)'s throughput. The acceptance probability decreases with more iterations.}


\begin{figure*}
	\centering
	\subfloat[Img-to-img\label{fig:1}]{
		\includegraphics[width=0.48\columnwidth,height=3cm]{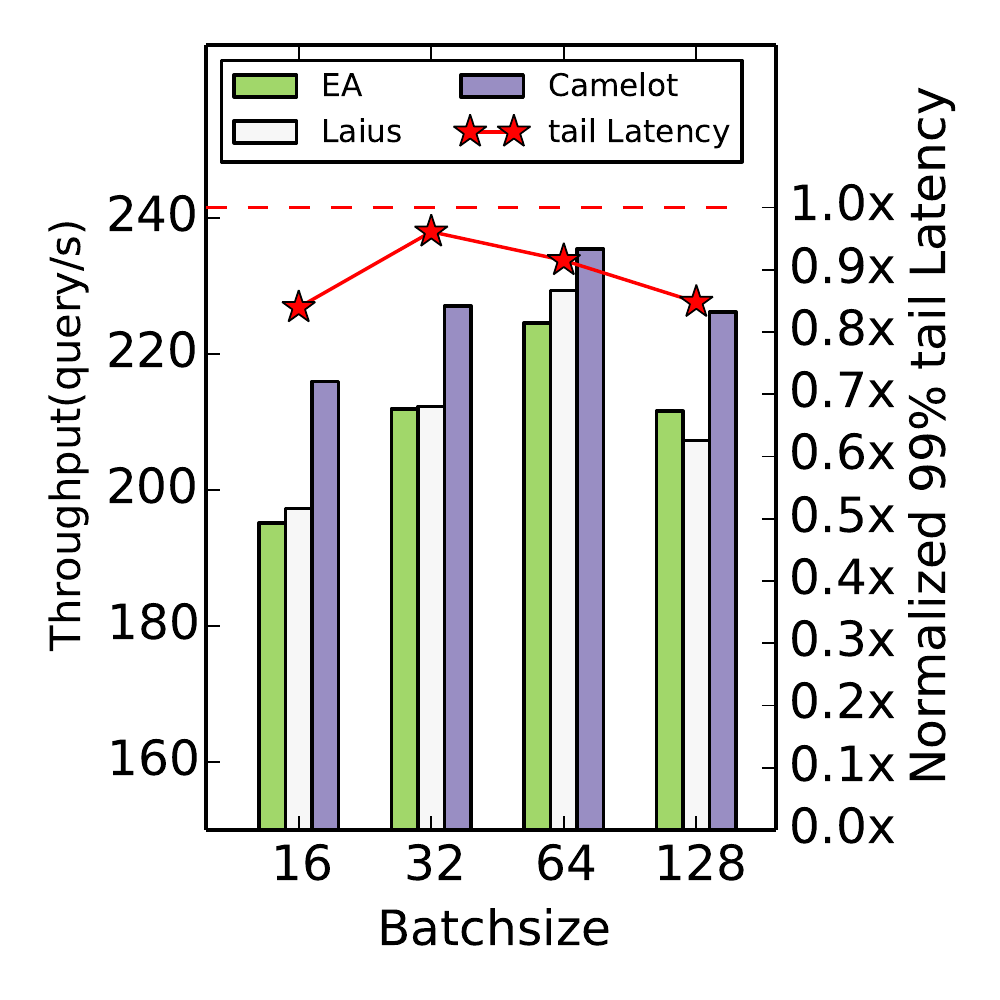}
	}
	\hfill
	\subfloat[Img-to-text\label{fig:2}]{
		\includegraphics[width=0.48\columnwidth,height=3cm]{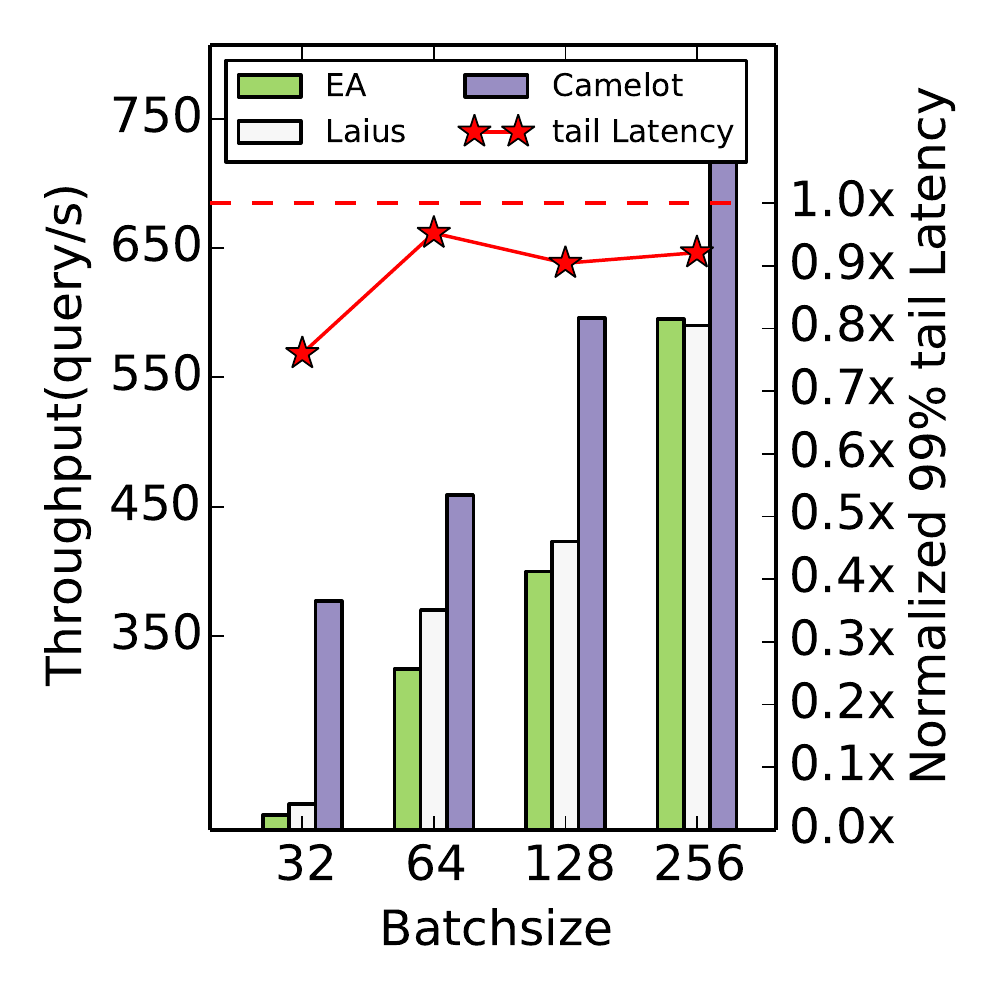}
	}
	\hfill
	\subfloat[Text-to-img\label{fig:3}]{
		\includegraphics[width=0.48\columnwidth,height=3cm]{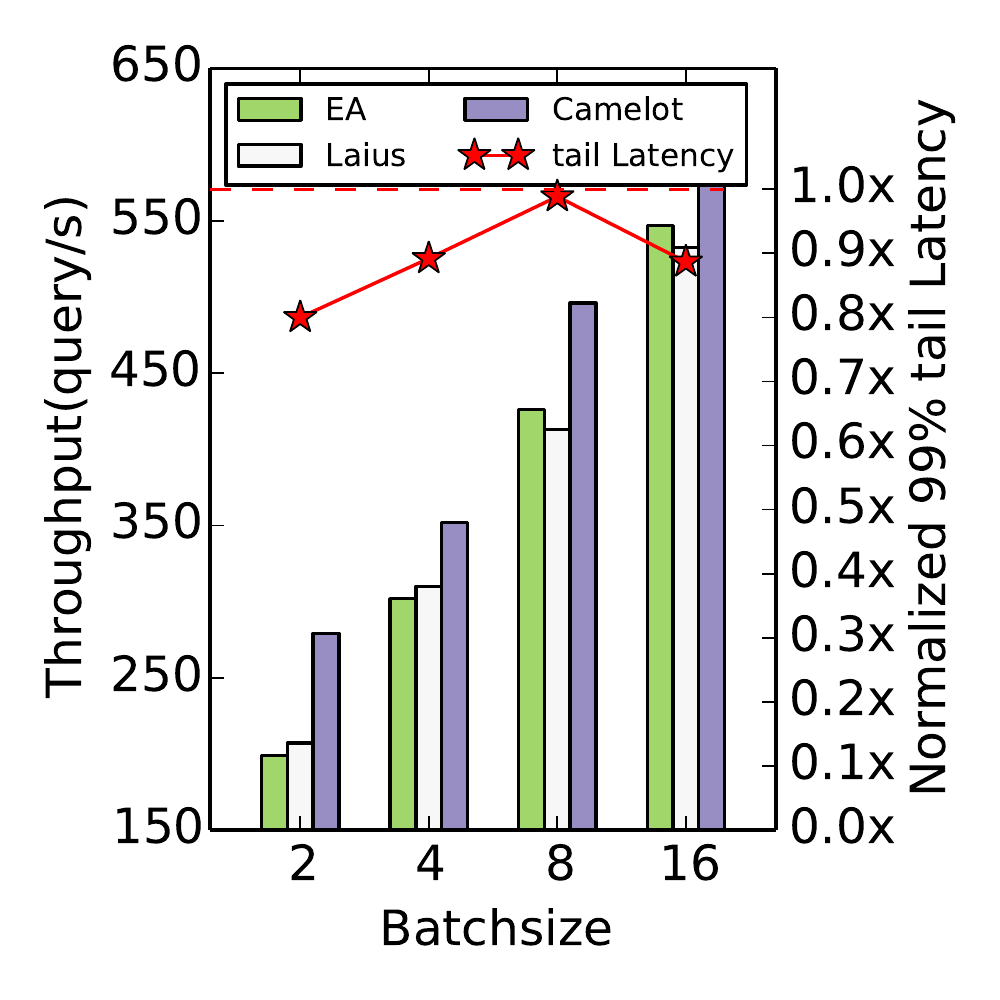}
	}
	\hfill
	\subfloat[Text-to-text\label{fig:4}]{
		\includegraphics[width=0.48\columnwidth,height=3cm]{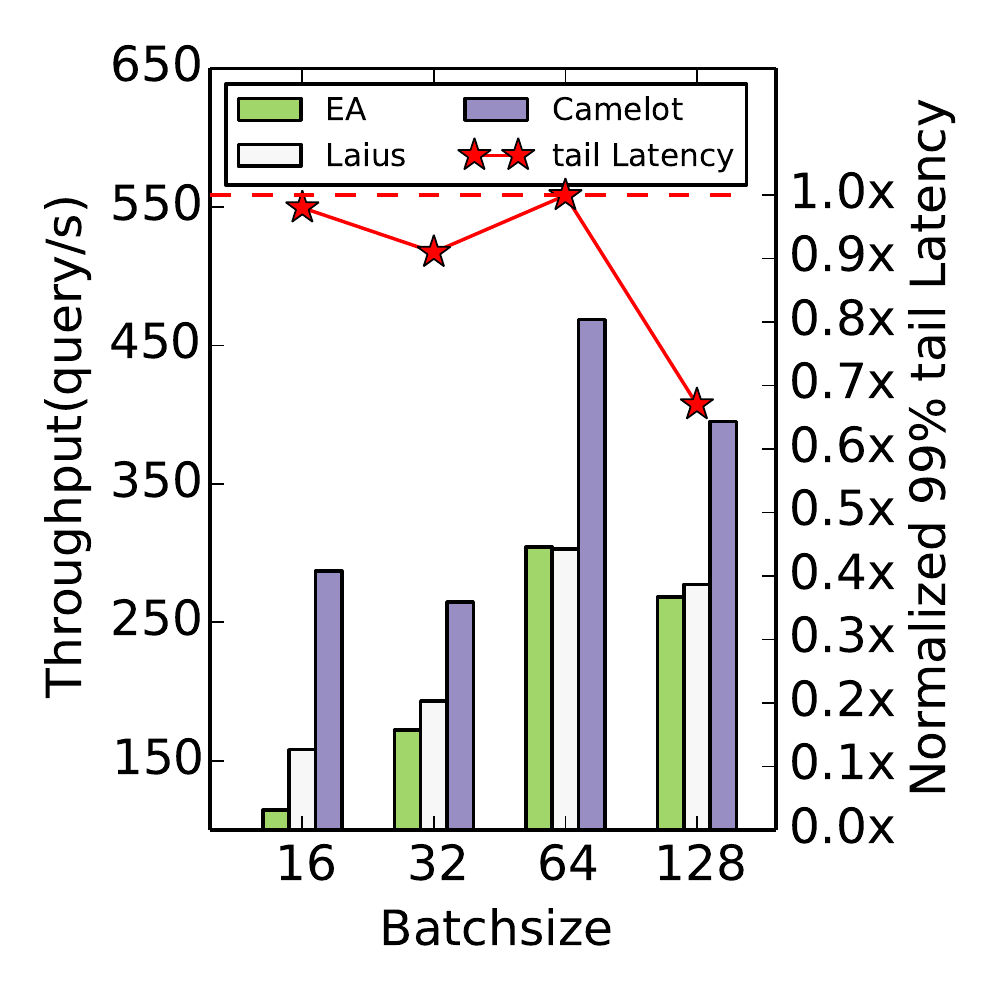}
	}
	\vspace{-2mm}
	\caption{\label{fig:realresult} The supported peak loads of the benchmarks with EA, Laius and Camelot. The stars shows the normalized 99\%-ile latencies of the benchmarks with Camelot (corresponding to the right $y$-axis). 
	}
		\vspace{-4mm}
\end{figure*}

\subsection{Deployment scheme across multiple GPUs}

Distributing microservice instances to multiple GPUs contains two steps.
The first step is also searching for the number of instances for each 
microservice stage and the computing resource quotas allocated to 
each instance. The second step is to find a deployment scheme according to 
the number of instances for each stage and the computing resource 
quotas in the first step. However, it is impractical to search 
exhaustively for the optimal deployment scheme for all instances. To 
speed up the entire search progress, we use a specific deployment 
strategy to quickly find out a reasonable deployment scheme as shown in
Figure~\ref{fig:deployment}.

A GPU has multiple resource dimensions including
computing resource, global memory capacity, global memory bandwidth and PCIe bandwidth, etc. When 
deploying the instances of a microservice stage, we sort the remaining 
GPUs according to their available resources. The partial ordering of 
resources during GPU sorting is related to the characteristics of the 
microservice. According to previous experiments in Section~\ref{sec:background}, we prove that for GPU microservices, the global memory capacity will become the major resource bottleneck. Therefore, Camelot sets the global memory capacity as the highest priority resource in the deployment scheme. For example, for applications that take up a lot of
global memory space, they will be sorted according to the size of the remaining 
global memory when sorting. If the remaining global memory is the same size, then 
they will be sorted according to other resource dimensions.

GPUs with fewer resources have higher priority and will try to deploy 
instances on the GPU with higher priority first. In this case, Camelot 
avoids excessive fragmentation of the resources available in the 
resource pool. In addition, deploying instances of
the same stage on the same GPU as much as possible can share models 
between multiple instances, reducing the consumption of GPU global
memory, which is often the most stressful resource during allocation.

\section{Evaluation of Camelot} \label{sec:evaluation}
In this section, we evaluate the effectiveness of Camelot in maximizing the supported peak load and minimizing resource usage at low load, while ensuring the required QoS.

We evaluate Camelot on a machine equipped with two Nvidia RTX 2080Ti GPUs
and a DGX-2 machine~\cite{dgx2} that equipped with Nvidia V100 GPUs.
Table~\ref{table:Specification} summarizes the detailed software and hardware experimental configurations. Camelot does not rely on any special hardware features of 2080Ti or V100, and is easy to be set up on other GPUs with Volta or Turing architecture. The peak global memory bandwidths of the 2080Ti and V100 GPUs are 616 GB/s and 897 GB/s, respectively~\cite{v100}. They are used as constraints in the resource allocation policies.
We use both the real system benchmarks and the artifact benchmarks in Camelot suite as user-facing GPU microservices. 
Except the large scale evaluation in Section~\ref{sec:general}, we report the experimental results on the machine equipped with two 2080Ti GPUs.
\begin{table}
	\caption{Hardware and software specifications.}
	\label{table:Specification}
	\scriptsize
	\centering
	\begin{tabular}{c|c}
		\hline
		\ & \textbf{Specification}\\
		\hline
		\multirow{4}*{\textbf{Hardware}} & Intel(R) Xeon(R) CPU E5-2620 v3 @ 2.40GHz \\
		~ & Two Nvidia GeForce RTX 2080Ti \\ 
		\cline{2-2}
		~ & Intel(R) Xeon(R) Platinum 8168 CPU @ 2.70GHz \\ 
		~ & NVIDIA DGX-2 with 16 Tesla V100s-SXM3\\
		\hline
		\multirow{2}*{\textbf{Software}} & Ubuntu 16.04.5 LTS with kernel 4.15.0-43-generic \\
		~ & CUDA Driver 410.78 CUDA SDK 10.0 CUDNN 7.4.2	\\
		\hline
	\end{tabular}
	\vspace{-3mm}
\end{table} 

While we do not find prior work on resource management for GPU microservices, we compare Camelot with the {\it Even allocation} (``EA'' for short) policy, and Laius~\cite{zhang2019laius} that is proposed for managing the applications co-located on spatial multitasking GPUs. EA evenly allocates all the GPU resources to the microservices in a user-facing applications. On a spatial multitasking GPU, Laius predicts the computational resource required by a user-facing query and dynamically reallocates the remaining computational resources to batch applications for maximizing their throughputs. 
While Laius is designed for single GPU situation, we schedule the microservices of a benchmark on a single GPU with Laius. The total throughput of the benchmark  with Laius is calculated by aggregating the throughputs on all the GPUs. 

\subsection{Maximizing the Supported Peak Load}
\label{sec:maximize}
In this subsection, we evaluate Camelot in maximizing the supported peak load while ensuring the required QoS with a given number of GPUs. 


Figure~\ref{fig:realresult} shows the supported peak loads of the benchmarks normalized to their QoS targets with EA, Laius and Camelot, while ensuring the 99\%-ile latency target. In the figure, the $x$-axis shows the batch size of processing user queries.
Camelot increases the supported peak loads of the benchmarks by 12\% to 73.9\% compared with EA, and by 10\% to 64.5\% compared with Laius. 

{\color{black}
EA results in the low peak loads of the benchmarks because it does not consider the pipeline effect of the microservices. While the peak load of a benchmark is determined by the the peak load of the microservice stage that shows the lowest throughput, the resource allocation does not balance the throughputs of the microservice stages. In addition, the benchmarks achieve slightly higher peak load with Laius compared with EA. This is mainly because we already optimize Laius to balance the throughputs of the microservice stages. However, Laius still performs worse than Camelot, because it does not schedule microservice instances across multiple GPUs as Camelot does. In this case, the microservices suffer from higher contention with Laius compared with Camelot. In addition, the benchmarks suffer from the long communication overhead without the global memory-based communication in EA and Laius.
}

\begin{figure}[!t]
	\vspace{-2.5mm}
	\subfloat[The number of instances for each microservice stage\label{fig:instance}]{
		\includegraphics[width=0.47\columnwidth,height=3cm]{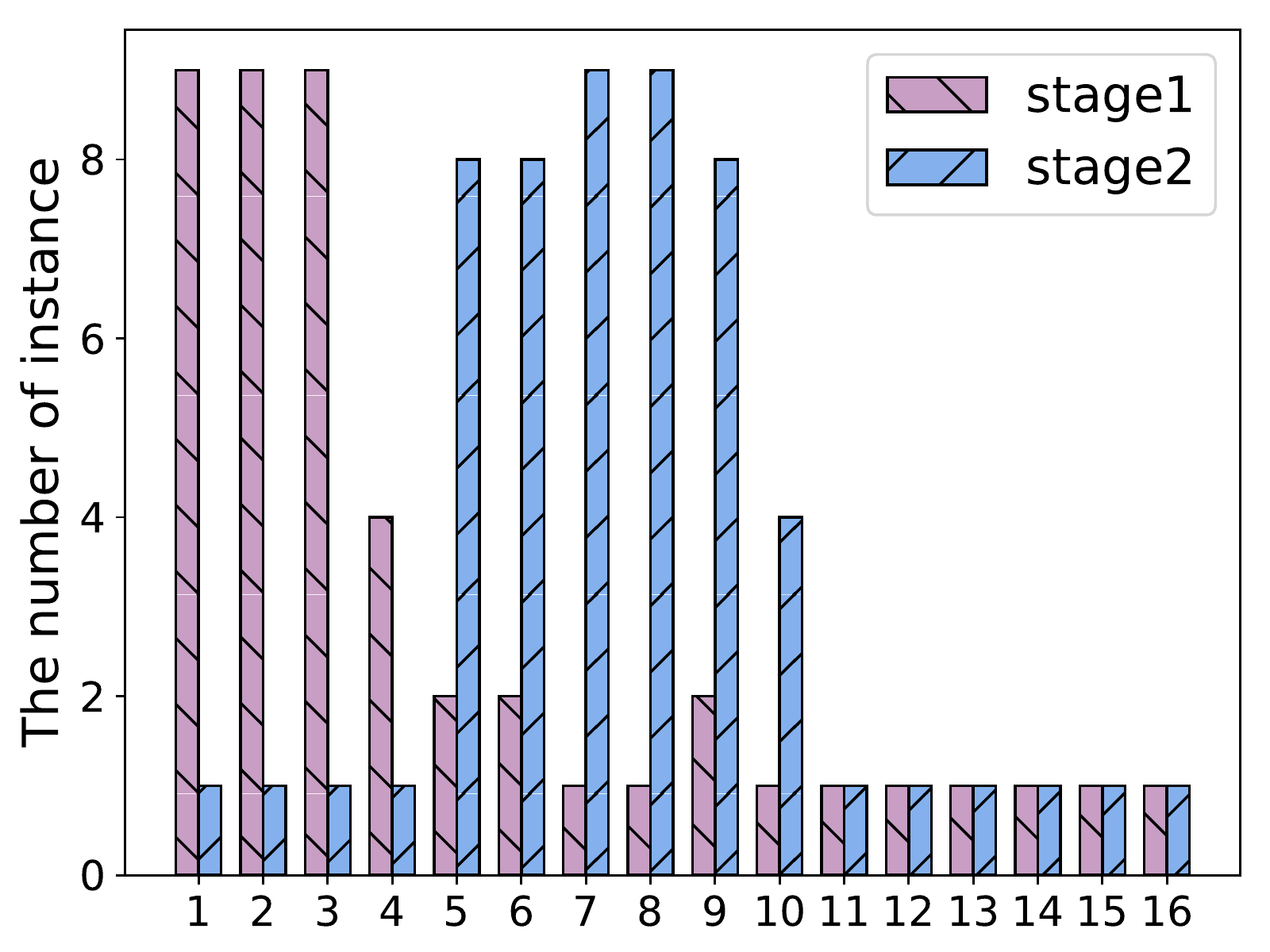}
	}
	\hfill
	\subfloat[The percentage of SMs allocated to each microservice\label{fig:resourcesquota}]{
		\includegraphics[width=0.47\columnwidth]{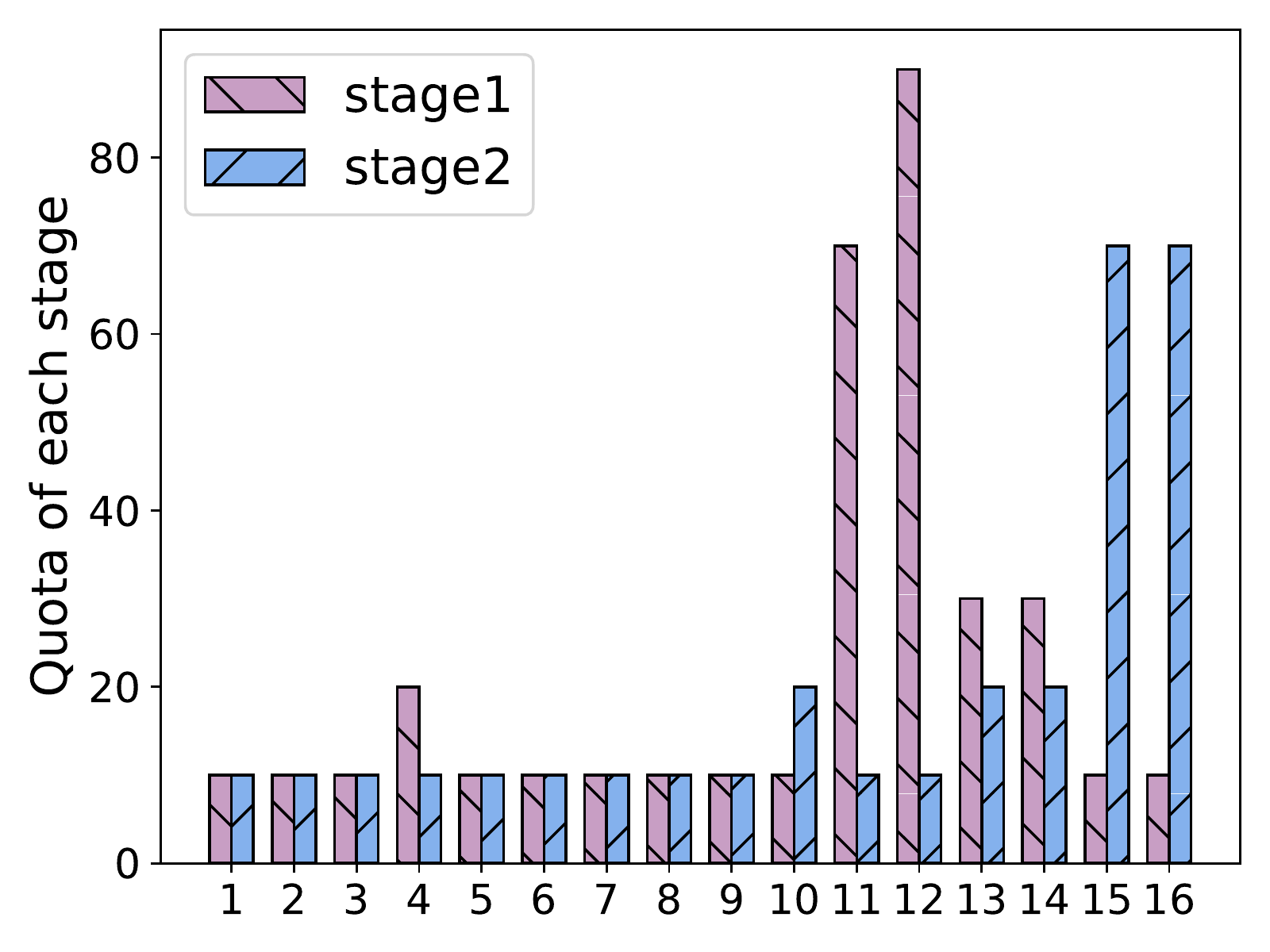}
	}
		\vspace{-2mm}
\caption{\label{fig:realbench-breakdown}The detailed resource allocation with Camelot.}
	\vspace{-4mm}
\end{figure}

In more detail, Figure~\ref{fig:realbench-breakdown} shows the number of instances in each microservice stage, and the percentage of SMs allocated to each microservice instance with Camelot.
The 16 test cases in Figure~\ref{fig:realresult} are referenced to be 1-16 for simplicity in Figure~\ref{fig:realbench-breakdown}. Observed from this figure, for the microservice stage that has long processing time (e.g., stage 1 for {\it img-to-img}), Camelot automatically creates more instances for it to  increase its total throughput. In the way, Camelot improves the pipeline efficiency of GPU microservices. 

\subsection{Minimizing Resource Usage}
\label{sec:minimize}

\begin{figure}
	\centering
	\includegraphics[width=.8\columnwidth,height=3cm]{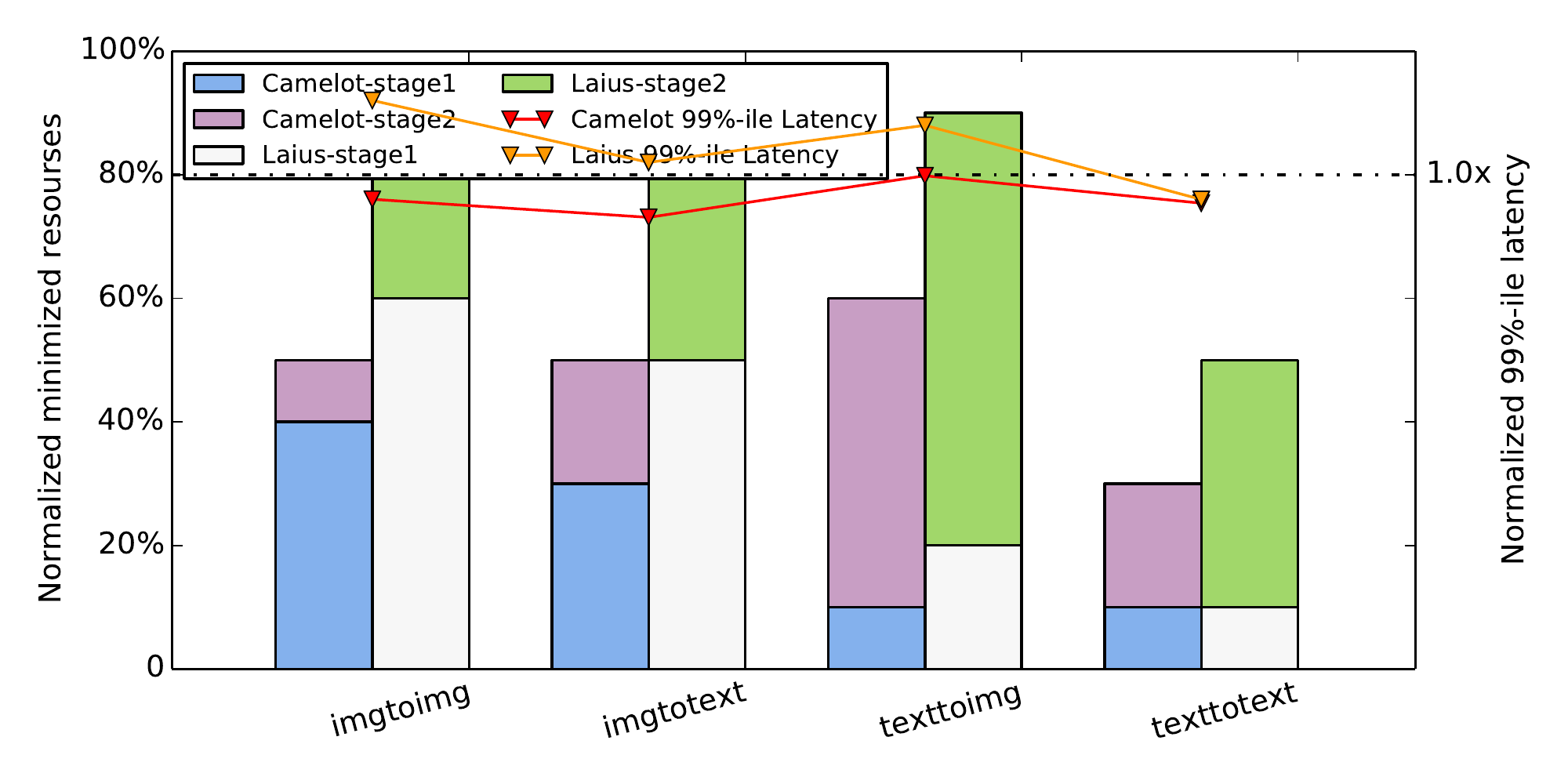}
	\vspace{-4mm}
	\caption{\label{fig:minimize} The efficiency of Camelot and Laius in reducing the resource usage.}
		\vspace{-4mm}
\end{figure}

Figure~\ref{fig:minimize} shows the normalized GPU resource usage of the benchmarks  at low load and the corresponding 99\%-ile latency with Camelot and Laius. We choose to use 30\% of the peak load to be the low load in the experiment as reported by Google's research~\cite{barroso2009datacenter}. In this figure, the resource usage is normalized to the scenario that each microservice stage uses an individual GPU. The expeirment with other loads show similar result. 

Observed from Figure~\ref{fig:minimize}, Camelot reduces the GPU resource usage 
by 46.5\% on average while ensuring the QoS of all the benchmarks. 
Camelot is effective in this scenario because it precisely predicts the 
duration of a microservice with different GPU resource configurations, and schedules 
microservice instances considering the runtime shared resource contention (global memory
bandwidth, and PCIe bandwidth). 
{\color{black}
Laius also reduces the resource usage compared with the naive deployment by 20.2\% on average.
However, because it does not optimize the inter-microservice contention and does not 
adjust the number of instances for each microservice stage, it requires more resource 
than Camelot to ensure the QoS of user-facing applications. Camelot reduces the GPU resource usage by 35\%, 
while Laius results in slight QoS violation for 3 out of the 4 benchmarks.
}

\begin{figure}
	\centering
	\includegraphics[width=.8\columnwidth]{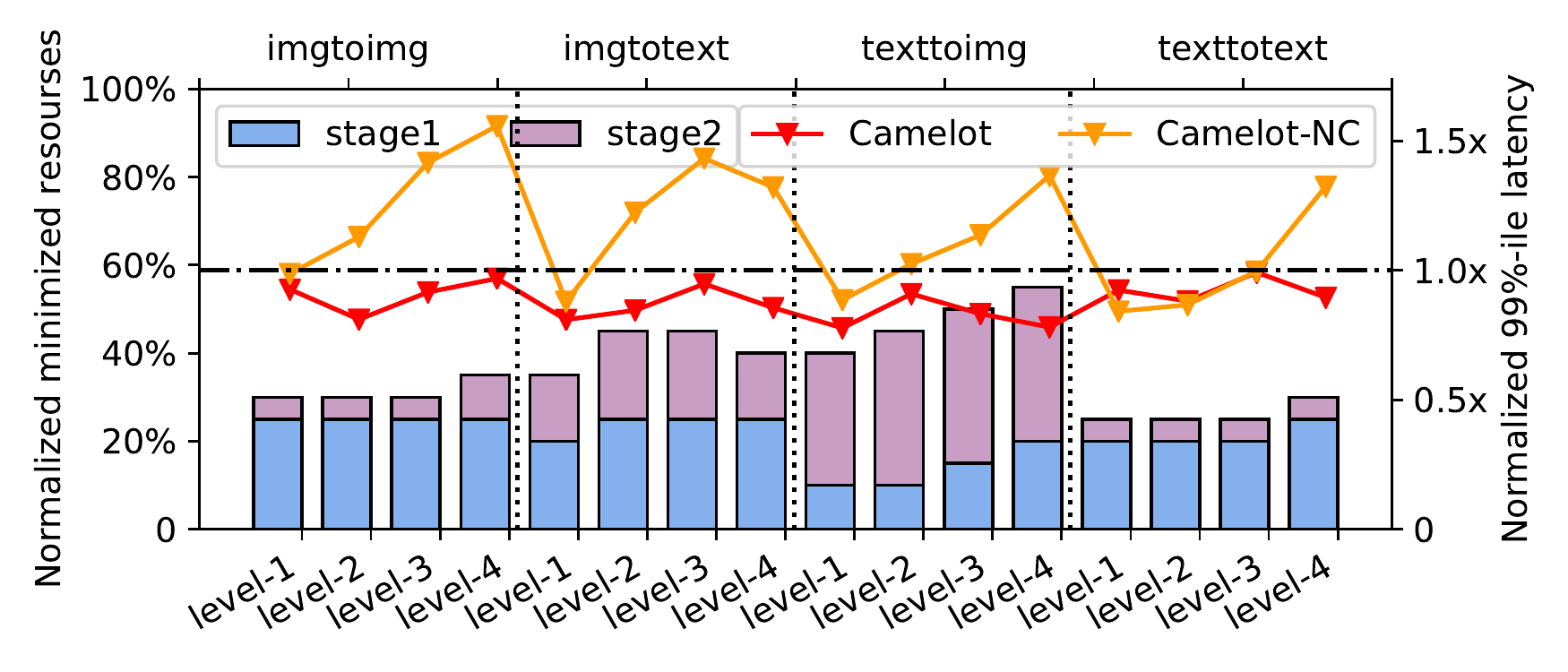}
	\vspace{-4mm}
	\caption{\label{fig:loadchange} The resource usages of the benchmarks under different load levels with Camelot, and the 99\%-ile latencies of the benchmarks with Camelot, and Camelot-NC.}
		\vspace{-4mm}
\end{figure}

\subsection{Adapting to Different Loads}
In this subsection, we evaluate Camelot in adapting the different loads.
For each benchmark, we report its resource usages and the corresponding 99\%-ile latencies under four different loads with Camelot in Figure~\ref{fig:loadchange}.
In the figure, the load of level $i$ is higher than the load of level $j$, if $i>j$.
\begin{figure}
	\centering
		\vspace{-5mm}
	\includegraphics[width=.9\columnwidth]{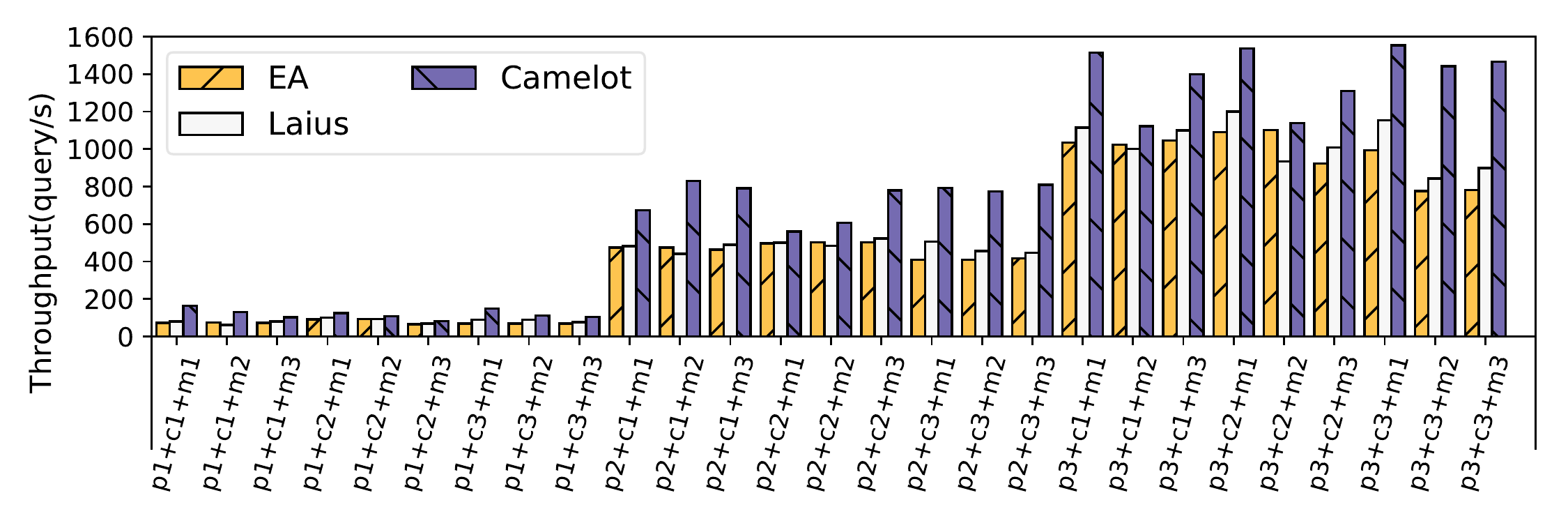}
	\vspace{-3mm}
	\caption{\label{fig:result1}The throughputs of the artifact benchmarks with EA, Laius, and Camelot.}
	\vspace{-3mm}
\end{figure}

\begin{figure*}
	\centering
	\subfloat[Img-to-img\label{fig:1}]{
		\includegraphics[width=0.47\columnwidth,height=3cm]{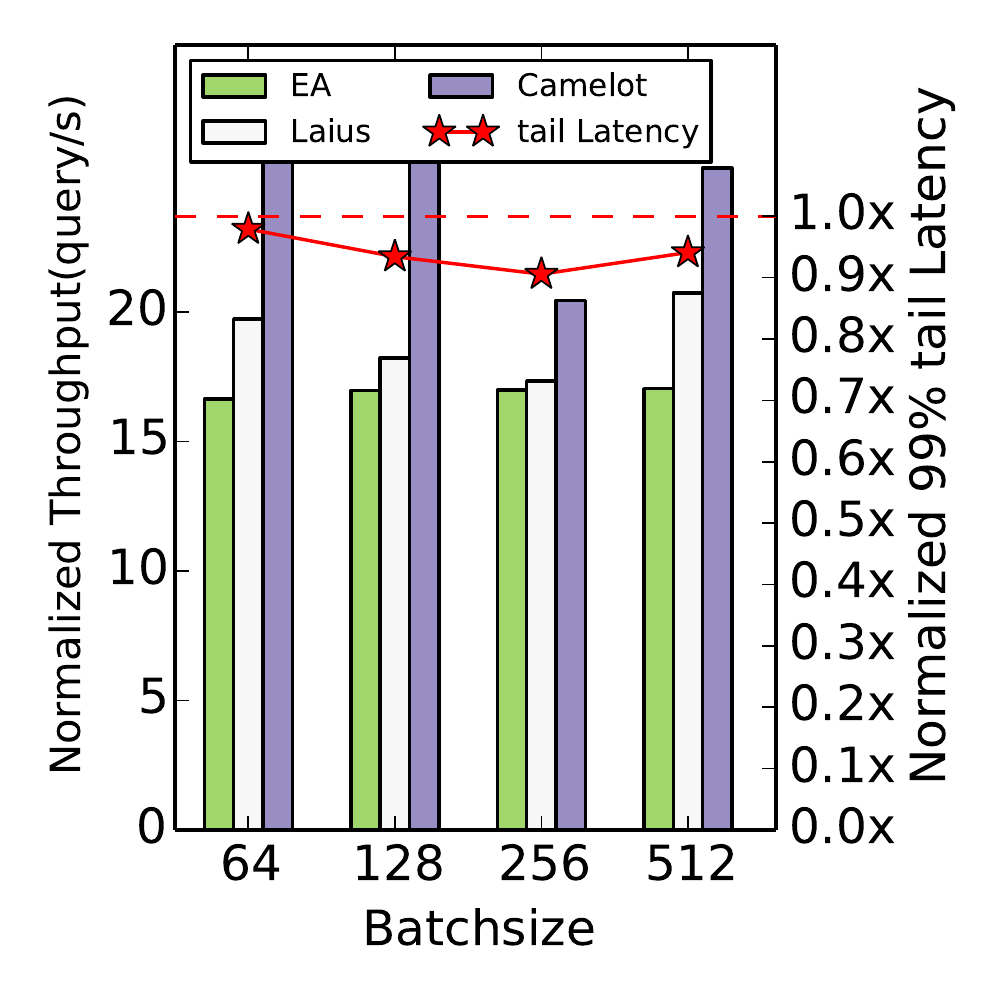}
	}
	\hfill
	\subfloat[Img-to-text\label{fig:2}]{
		\includegraphics[width=0.47\columnwidth,height=3cm]{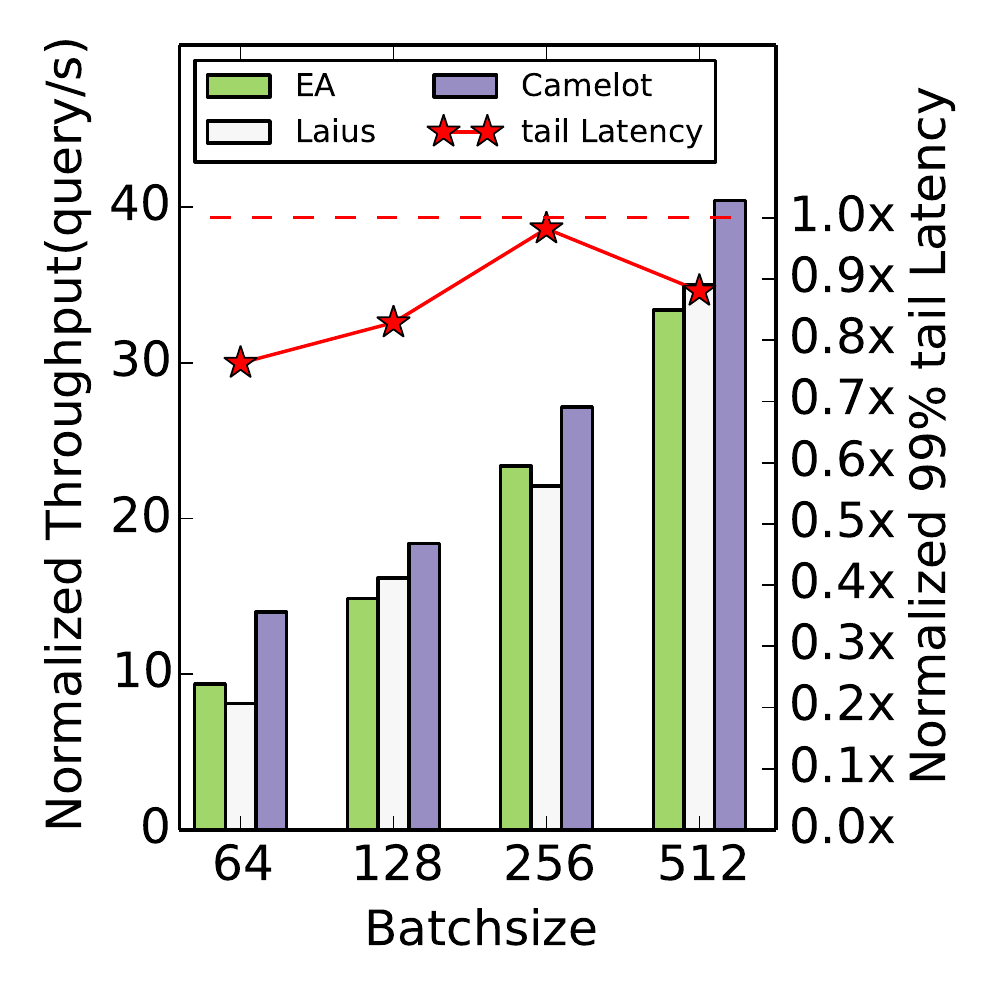}
	}
	\hfill
	\subfloat[Text-to-img\label{fig:3}]{
		\includegraphics[width=0.47\columnwidth,height=3cm]{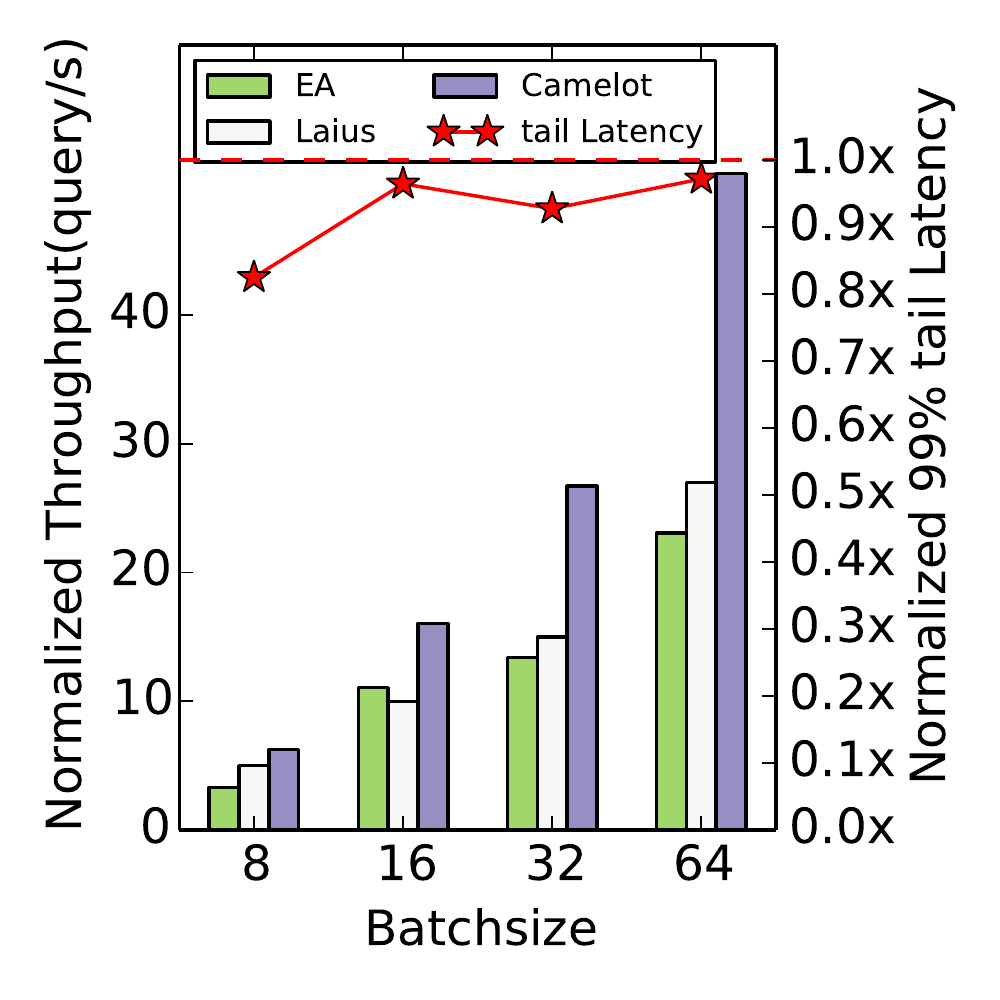}
	}
	\hfill
	\subfloat[Text-to-text\label{fig:4}]{
		\includegraphics[width=0.47\columnwidth,height=3cm]{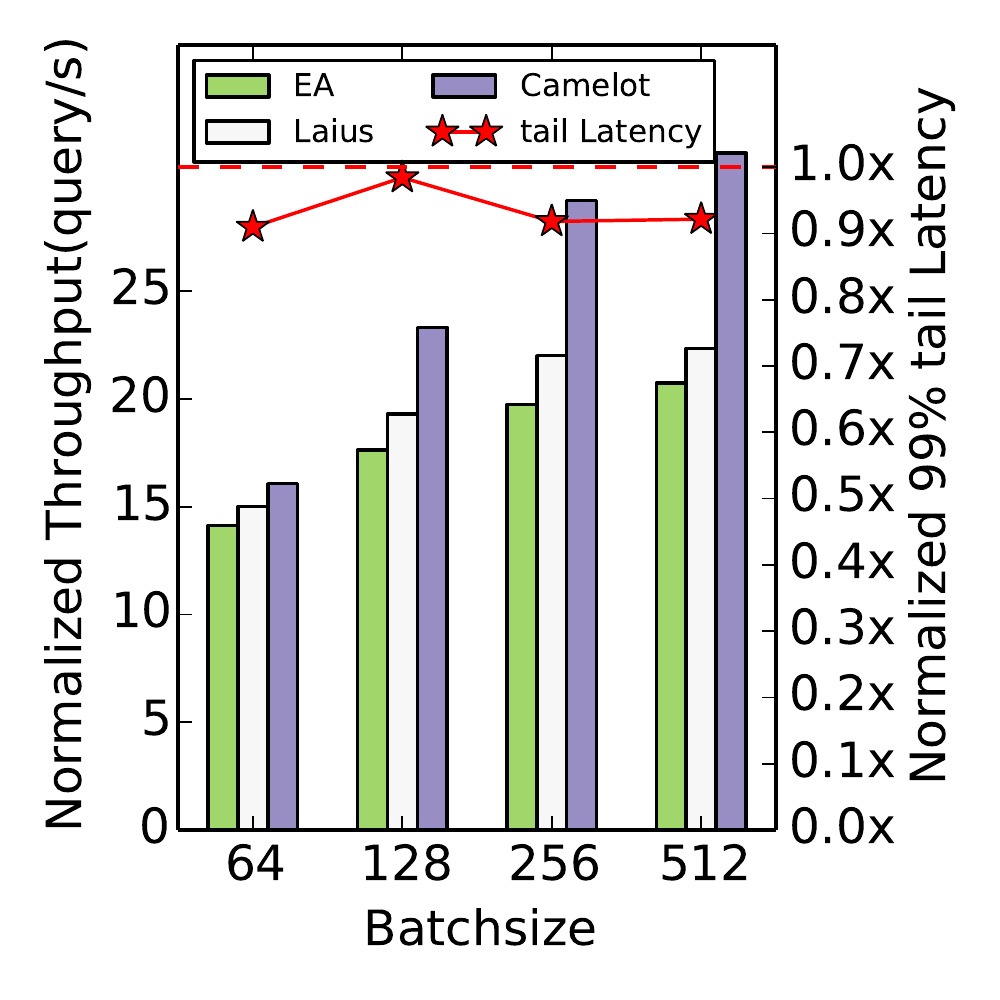}
	}
	\vspace{-2mm}
	\caption{\label{fig:DGX-2-max} The supported peak loads of the benchmarks on DGX-2 with EA, Laius and Camelot. The stars shows the normalized 99\%-ile latencies of the benchmarks with Camelot (corresponding to the right $y$-axis). 
	}
		\vspace{-3mm}
\end{figure*}

Observed from this figure, Camelot reduces more resource usage when the load is lower, and always guarantee the QoS of the benchmarks. 
Camelot is able to fine tune the GPU resource allocation based on the load, and the contention between the microservices on the same GPU.


\subsection{Effectiveness of Constraining Global Memory Bandwidth Contention}
\label{sec:eval-contention}
Camelot predicts the global memory bandwidth usage of all the microservices, and makes sure that the accumulated bandwidth usage of the concurrent tasks is smaller than the peak global memory bandwidth of the GPU. 
To show the effectiveness of this constraint, we implement Camelot-NC, a system that disables the constraint in Camelot.

Figure~\ref{fig:loadchange} also shows the the 99\%-ile latency of the benchmarks with Camelot-NC.
 Observed from this figure, user-facing services in 10 out of the 16 test cases suffer from QoS violation with Camelot-NC. For instance, the 99\%-ile latency of {\it img-to-img} is up to 1.55X of its QoS target with Camelot-NC. The QoS violation is due to the unmanaged global memory bandwidth contention.



\subsection{Generalizing for Complex Microservices}
\label{sec:general}
Besides the real-system benchmarks, we create $3\times3\times3 = 27$ more benchmarks using the artifact benchmark in Camelot suite (3 microservices with different compute intensities, 3 microservices with different memory access intensities, and 3 microservices with different PCIe intensities) to evaluate Camelot for complex microservices. The microservices are denoted by $c_1$, $c_2$, $c_3$, $m_1$, $m_2$, $m_3$, $p_1$, $p_2$, and $p_3$ respectively.
c$_i$/m$_i$/p$_i$ is more PCIe/compute/memory intensive than c$_j$/m$_j$/p$_j$, if $i>j$. 

Figure~\ref{fig:result1} shows the supported peak loads of the 27 artifact benchmarks with EA, Laius, and Camelot. In the figure, ``$p_i$+$c_i$+$m_i$''represents a benchmark that is built by pipelining a PCIe-instensive microservice $p_i$, a compute-intensive microservice $c_i$ and a memory-instenvie microservice $m_i$. Observed from this figure, 
on average, Camelot improves the supported peak load of the 27 benchmarks by 44.91\% compared to EA, and by 39.72\% compared with Laius.
\begin{figure}
	\centering
	\includegraphics[width=.8\columnwidth]{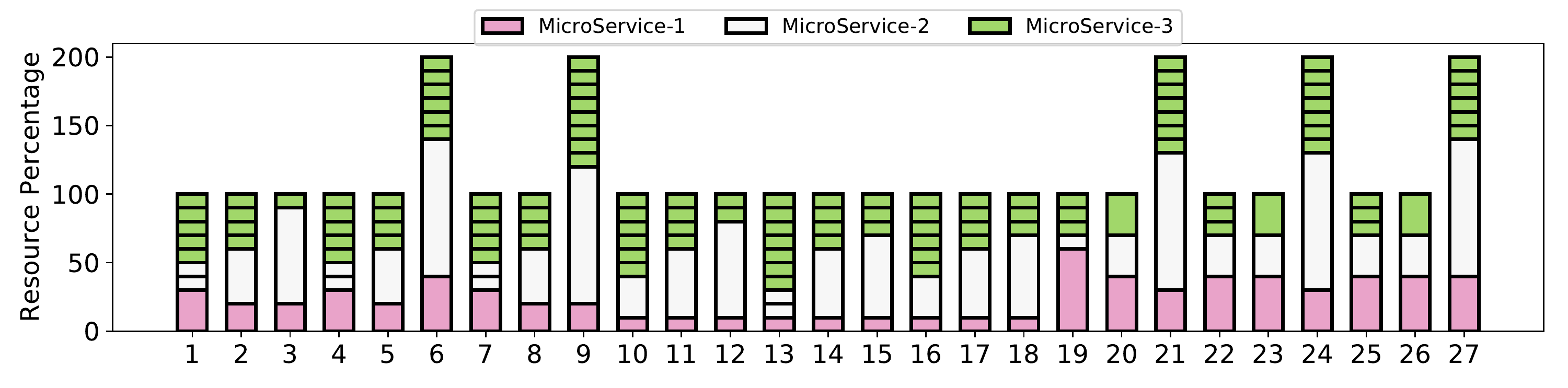}
	\vspace{-3mm}
	\caption{\label{fig:qos2} Resource allocation for maximizing the peak supported load of the benchmarks with Camelot.}
	\vspace{-3mm}
\end{figure} 

\begin{figure}
	\centering
	\includegraphics[width=.8\columnwidth]{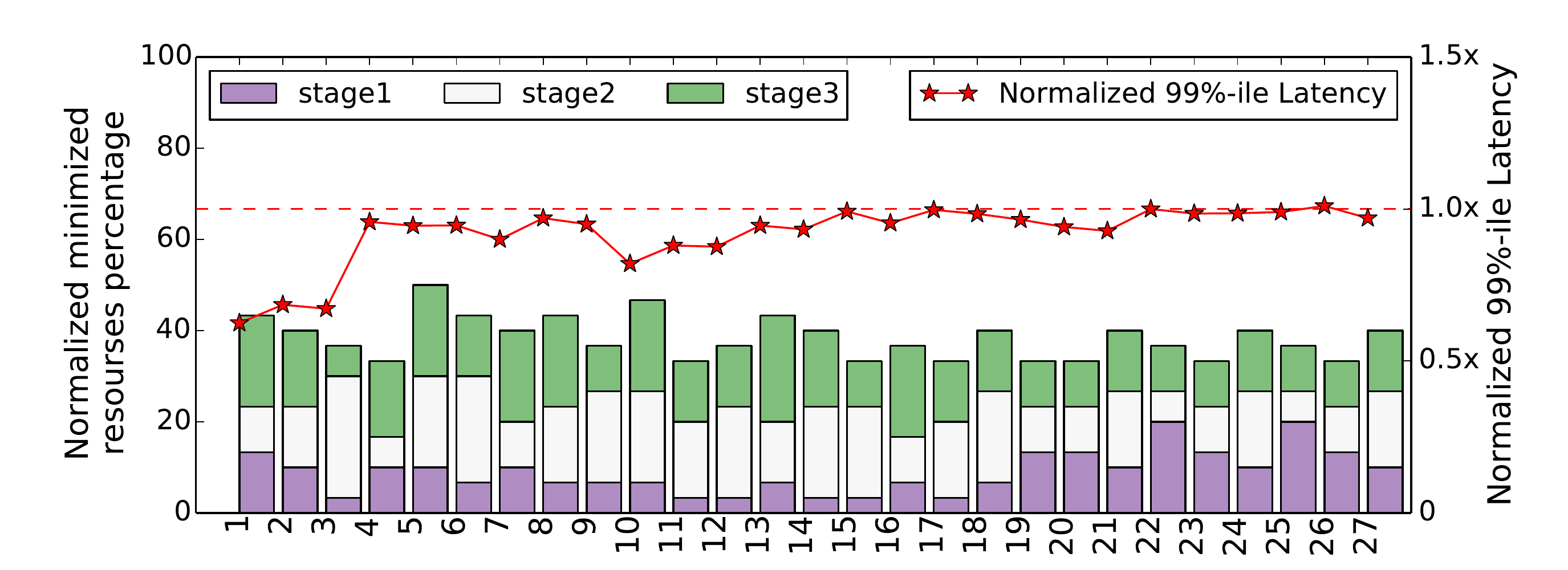}
	\vspace{-3mm}
	\caption{\label{fig:minimize-artifact} Resource allocation for the benchmarks at low loads and the corresponding 99\%-ile latencies with Camelot.}
		\vspace{-6mm}
\end{figure}

Corresponding to Figure~\ref{fig:result1}, Figure~\ref{fig:qos2} shows the resource allocation with Camelot for the 27 benchmarks.
Observed from this figure, Camelot launches different numbers of instances for different microservice stages, and allocates different percentages of the SMs to the microservices. For instance, Camelot launches 1 instance of Microservice-1, 2 instances of Microservice-2, and 5 instances of Microservice-3 in the first benchmark. In addition, Camelot allocates different percentages of the SMs to the same microservice when it is linked in different benchmarks. It reveals that Camelot is able to automatically adjust the resource allocation based on the features of the microservices.

Figure~\ref{fig:minimize-artifact} shows the resource usages and the corresponding 99\%-ile latencies of the 27 benchmarks at low 
load  with Camelot. 
Camelot significantly reduces the resource usage by 61.6\% on average. 
In addition, the GPU resource allocations vary for the 27 benchmarks.
This is because Camelot adjusts the resource allocation based on the characters of the pipelined microservices.
To conclude, Camelot is generalizable for complex microservices.

\subsection{Large Scale Evaluation on DGX-2}
We also evaluate Camelot on a large-scale DGX-2 machine in 
maximizing the supported peak load. We do not show the result of minimizing the
resource usage here because it is the same to the one on RTX 2080Ti.

Figure~\ref{fig:DGX-2-max} shows the supported peak loads of the benchmarks normalized to their QoS targets with EA, Laius and Camelot, while ensuring the 99\%-ile latency target.
In the figure, the $x$-axis shows the batch sizes of processing user queries.
Observed from this figure, Camelot increases the supported peak load by 50.1\% for all the benchmarks on average compared with EA, while guaranteeing their 99\%-ile latency within the required QoS target. 
{\it Camelot is scalable on large-scale GPU machines}.

\subsection{Overhead of Camelot}
{\bf Offline overhead}. The overhead of training models offline for predicting microservice performance is acceptable. We collect the training samples of all the microservices within a single day using a single GPU. We can further speed up the sample collection by using multiple GPUs. As for the online predicting, each prediction completes in 1 ms, which is much shorter than the QoS target of a service. 
{\bf Resource allocation overhead}. As stated in Section~\ref{sec:schedule}, Camelot needs to solve the optimization problem using the simulated annealing algorithm to identify the appropriate resource allocation. {\color{black} Our measurement shows that this operation completes in 5ms}. 
{\bf Communication overhead}. Camelot need to setup global memory-based communication for microservices that require data transfer. The setup operation based on CUDA IPC technique for a pair of microservices is only done once when the end-to-end service is launched. The setup operation completes in 1ms.
To conclude, the overhead of Camelot is acceptable for real-system deployment. 


\section{Conclusion} \label{sec:conclusion}
For GPU microservices, the main memory-based communication between 
the microservices, the pipeline inefficiency, and the global memory bandwidth contention result 
in their poor performance. To this end, we propose 
Camelot, a runtime system to manage GPU resources online.
Camelot uses a global memory-based communication mechanism to eliminate the large communication overhead.
We also propose 
two contention-aware resource allocation policies
that considers the pipeline efficiency and shared resource contention.
Experimental results show that Camelot increases the peak supported load by up to 
64.5\%, and reduces 35\% resource usage at low load while achieving the desired 99\%-ile latency target compared with the state-of-the-art work.

\bibliographystyle{IEEEtran}
\bibliography{refs}

\end{document}